\documentclass[acmsmall]{acmart}
\usepackage[caption=false,font=footnotesize]{subfig}
\usepackage{caption}
\usepackage{multirow}
\usepackage{makecell}
\usepackage{bm}
\AtBeginDocument{%
  \providecommand\BibTeX{{%
    \normalfont B\kern-0.5em{\scshape i\kern-0.25em b}\kern-0.8em\TeX}}}

\setcopyright{acmcopyright}
\copyrightyear{2018}
\acmYear{2018}
\acmDOI{10.1145/1122445.1122456}

\acmJournal{JACM}
\acmVolume{37}
\acmNumber{4}
\acmArticle{111}
\acmMonth{8}

\begin{document}

\title{A Multimodal Framework for Large-Scale Emotion Recognition by Fusing Music and Electrodermal Activity Signals}

\author{Guanghao Yin}
\email{ygh_zju@zju.edu.cn}
\affiliation{%
  \institution{Zhejiang University}
  \streetaddress{Zheda Road 38}
  \city{Hangzhou}
  \country{China}
  \postcode{43017-6221}
}

\author{Shouqian Sun}
\email{ssq@zju.edu.cn}
\affiliation{%
  \institution{Zhejiang University}
  \streetaddress{Zheda Road 38}
  \city{Hangzhou}
  \country{China}
}

\author{Dian Yu}
\email{yudian329@zju.edu.cn}
\affiliation{%
  \institution{Zhejiang University}
  \streetaddress{Zheda Road 38}
  \city{Hangzhou}
  \country{China}
}

\author{Dejian Li}
\email{dejianli@zju.edu.cn}
\affiliation{%
  \institution{Zhejiang University}
  \streetaddress{Zheda Road 38}
  \city{Hangzhou}
  \country{China}
}

\author{Kejun Zhang}
\authornote{Corresponding author.}
\email{zhangkejun@zju.edu.cn}
\affiliation{%
  \institution{Zhejiang University and Alibaba-Zhejiang University Joint Research Institute of Frontier Technologies}
  \streetaddress{Zheda Road 38}
  \city{Hangzhou}
  \country{China}
}

\renewcommand{\shortauthors}{Yin and Zhang, et al.}

\begin{abstract}
  Considerable attention has been paid to physiological signal-based emotion recognition in the field of affective computing. For reliability and user-friendly acquisition, electrodermal activity (EDA) has a great advantage in practical applications. However, EDA-based emotion recognition with large-scale subjects is still a tough problem. The traditional well-designed classifiers with hand-crafted features produce poorer results because of their limited representation abilities. And the deep learning models with auto feature extraction suffer the overfitting drop-off because of large-scale individual differences. Since music has a strong correlation with human emotion, static music can be involved as the external benchmark to constrain various dynamic EDA signals. In this paper, we make an attempt by fusing the subject's individual EDA features and the external evoked music features. And we propose an end-to-end multimodal framework, the one-dimensional residual temporal and channel attention network (RTCAN-1D). For EDA features, the channel-temporal attention mechanism for EDA-based emotion recognition is first involved in mine the temporal and channel-wise dynamic and steady features. The comparisons with single EDA-based SOTA models on DEAP and AMIGOS datasets prove the effectiveness of RTCAN-1D to mine EDA features. For music features, we simply process the music signal with the open-source toolkit openSMILE to obtain external feature vectors. We conducted systematic and extensive evaluations. The experiments on the current largest music emotion dataset PMEmo validate that the fusion of EDA and music is a reliable and efficient solution for large-scale emotion recognition.
\end{abstract}



\maketitle

\section{Introduction}
The ability of emotion recognition is a significant hallmark of intelligent human-computer interaction (HCI). Empowering computers to discern human emotions better would make them perform the appropriate actions. In recent two decades, both the research groups and industries have aroused great interest that explores to enhance user experience with algorithms \cite{picard2000affective}. With the advancement of sensor-based Internet of Things (IoT) technology, emotion recognition is in unceasing demand by a wide range of applications such as mental healthcare \cite{guo2013pervasive}, driving monitor \cite{katsis2008toward}, wearable devices \cite{sano2013stress}, etc.

Human emotions are usually defined as a group of affective states that evoke stimuli from external environments or interpersonal events as a response \cite{picard2001toward}. According to behavioral researches, emotions are typically described as the set of discrete basic categories \cite{ekman1992argument}. They can also be represented in the continuous space \cite{hamann2012mapping}, which can quantitatively describe the abstract emotion state percentage, which provides another method for understanding the concrete representations of abstract emotions.
The valence-arousal space is one of the common emotion discrete spaces, where the valence reflects the value of the positive to negative state and the arousal measures the active to calm degree \cite{lang1993looking}. For the advantage of conducting the Self-Assessment Manikins (SAM), most of the popular open-source emotion datasets applied the valence-arousal annotations, such as DEAP \cite{koelstra2011deap}, AMIGOS \cite{correa2018amigos}, PMEmo \cite{zhang2018pmemo}, or etc.

In general, the data source of emotion recognition can be categorized into (1) peripheral physical signals such as facial expression \cite{kim2015emotion}, speech \cite{el2011survey}; (2) physiological signals including electroencephalography (EEG), electrodermal activity (EDA), electrocardiogram (ECG), or etc. The first method has great advancement in data collection. However, it cannot guarantee reliability in recognition because human can consciously control body language to hide their personal emotion-like speech or facial expression. Using physiological signals can overcome this drawback. The physiological signals are reacted from the autonomic and somatic nervous systems (ANS and SNS). They are largely involuntarily activated and cannot be triggered by any conscious or intentional control~\cite{jerritta2011physiological}. Thus, the physiological signal-based method provides an avenue to recognize affect changes that are less obvious to perceive visually \cite{jerritta2011physiological}. Compared with other physiological signals, the acquisition of EDA can be conveniently accomplished by intelligent wearable devices like smartwatches and wisebraves~\cite{anusha2018dry}. The research of EDA-based emotion recognition can be put faster into practical applications such as emotion monitor~\cite{nittala2020physioskin}, music liking recommendation~\cite{chiliguano2016hybrid}, or etc. Hence, the EDA signal has great significance of scientific value and practicability.

To mine useful features from physiological signals, two broad methods exist: (1) the traditional well-designed methods with hand-crafted features, and (2) the deep learning models with auto feature extraction. According to previous studies, a wide range of statistical features have been explored to seek predictive power for the classification of emotions, including time domain, frequency domain, and time-frequency domain features \cite{shukla2019feature}. However, the hand-crafted feature-based methods suffer limitations in the following aspects. First, the traditional methods such as SCL have limited representation ability to extract the complicated EDA features from large-scale subjects. Second, the design of hand-crafted feature strongly depends on prior knowledge of statistics and physiology in the small-scale data. Researchers should take more and more effective hand-crafted features into account for better performance. It is a great challenge for their feature selection method. Although, Shukla~\cite{shukla2019feature} has provided a systematic exploration of the feature selection method, it cannot guarantee the robustness with large-scale complicated physiological signals. Recently, the effectiveness of deep learning model has been proven in tackling emotion recognition \cite{2019A}. Moreover, the end-to-end DNN model can directly learn discriminative features from data. Therefore, to involve less prior knowledge and explore the general solution for large-scale emotion recognition, we choose the end-to-end deep learning model.

Compared with single subject, establishing the relationship between emotional state and the physiological signals with large-scale subjects is a challenging problem, because more subjects involve more complicated individual specificity.
For emotion recognition with one subject, previous user-dependent system~\cite{kim2008emotion} has achieved more than 90\% accuracy. However, when extending the subject from one to group, the accuracy of user-independent models decreased greatly \cite{jerritta2011physiological} because of the individual specificity. Moreover, our experimental results in Section~\ref{single EDA} have further supported this. Specifically, we use various hand-crafted classifier and CNN-based models to conduct incremental subject classification. In relatively small-scale datasets with dozens of subjects, such as DEAP (32 subjects), AMIGOS (40 subjects), and 1/10 PMEmo datasets (46 subjects), the models can build accurate relationships between EDA and emotion central states. However, when gradually increasing the subject number of PMEmo dataset to the maximum 457, the representation power of traditional classifiers with hand-crafted features is too weak to produce poor results, and the CNN-based deep learning model suffers overfitting. Since all those two mainstream methods have obvious defections, mining the single EDA single cannot further improve the performance for large-scale emotion recognition.

\begin{figure*}
\centering
\includegraphics[scale=0.4]{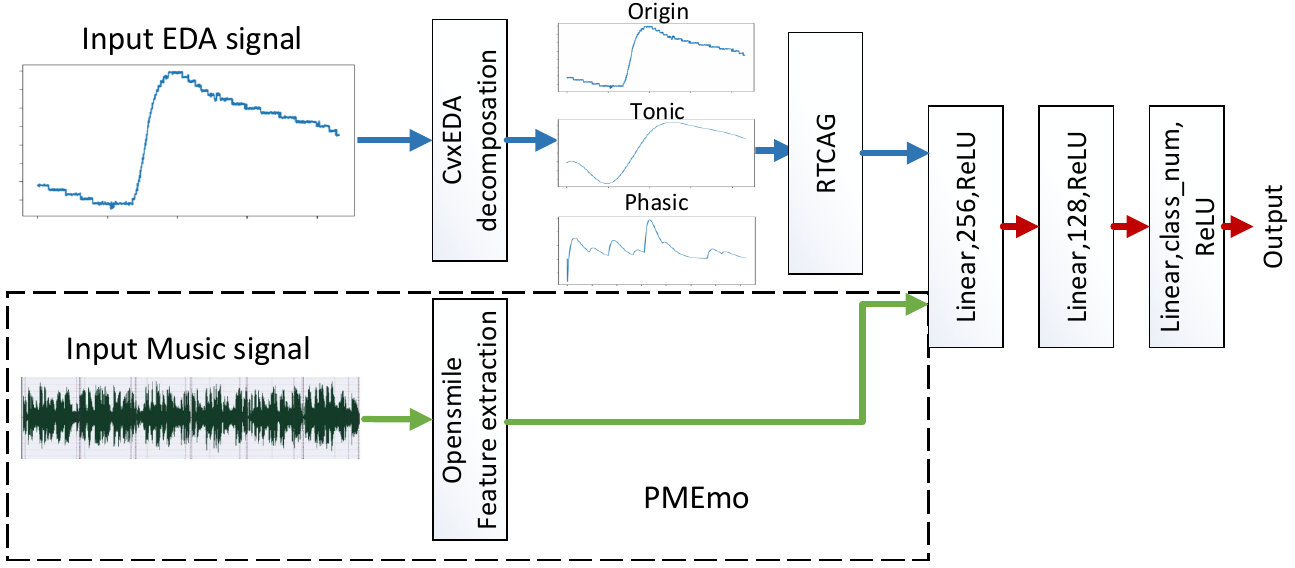}\\
\caption{The overall architecture of the proposed method. Music signals are only applied for PMEmo dataset, the music-evoked large-scale EDA dataset with hundreds of subjects. For the other three datasets, the model is only fed with single EDA input. For a fair comparison with previous works, the class number is set to 2 in DEAP, AMIGOS, and PMEmo. The structure of residual temporal and channel attention group (RTCAG) is illustrated in Fig. \ref{rtcag}.}
\label{architecture}
\end{figure*}

Towards this, we consider fusing external features with EDA signals. The multimodal fusion has the great advantage to enrich the complementary information from different modalities~\cite{baltruvsaitis2018multimodal}. It has a broad range of successes, including multi-sensor management~\cite{xiong2002multi}, and human-computer interaction~\cite{jaimes2007multimodal}. In the scene of psychological rehabilitation~\cite{langhorne2011stroke}, or emotion tracing~\cite{becker2004deep}, music is commonly utilized medium because it is effective to modulate subject' s mood \cite{lin2011exploiting}. In PMEmo dataset, music is what the authors use to induce emotions among participants. The tempo, beats, and other acoustic features regarding the music being played will correlate with the participants' observed emotions. Hence, we choose the music signal as the external constraint for supervised learning. As our experiments show, fusing the music features can efficiently improve the performance of RTCAN-1D and also overcome the defect that the EDA signal is relevant for arousal classification~\cite{torres2013feature,shukla2019feature}.

In this paper, we propose RTCAN-1D, a novel end-to-end 1D residual temporal and channel attention network for emotion recognition in the valence-arousal dimension. We directly apply a 1D residual block to build the EDA residual feature extraction (RFE). To provide sufficient useful messages from 1-channel EDA, we apply the convex optimization-based EDA method \cite{greco2015cvxeda} to decompose EDA into phasic, tonic components and feed the 3 channel mixed signals (origin, phasic, and tonic) into RTCAN-1D. Then, we involve the attention mechanism: (1) the signal channel attention module (SCA) to reweight the features of 3-mixed signal channels and (2) the residual nonlocal temporal attention module (RNTA) to explore the temporal correlations of EDA features. To the best of our knowledge, our model is the first attempt to involve the attention mechanism to process physiological signals. The EDA decomposition, two attention blocks, and residual feature extraction form the residual temporal and channel attention group (RTCAG) to get the EDA feature vectors. The external music features are simply extracted with the open-source toolkit openSMILE. Finally, the classifying layers fuse individual emotion features from EDA signals and external emotion benchmarks from music to predict the emotional state. The overall structure of our model is illustrated in Fig \ref{architecture} and Fig \ref{rtcag}.

We conduct extensive evaluations and compare our model with previous researches. Because the music and EDA signals from large-scale subjects are only available from the current largest PMEmo dataset, the multimodal fusion of RTCAN-1D is evaluated on PMEmo. Moreover, we prove the effectiveness of RTCAG with single EDA input from the small-scale DEAP and AMIGOS datasets. For a fair comparison with the SOTA methods, our proposed model was applied to the subject-independent binary classification of valence and arousal.

In summary, the contributions of our paper are represented as follows:
\begin{itemize}
\item An end-to-end multimodal framework RTCAN-1D is proposed for user-independent emotion recognition.
\item The effectiveness of the attention mechanism is first proved to capture the channel-wise correlations and long-distance temporal information of EDA signals.
\item The fusion of individual specificity from EDA signals and external static benchmarks from music provides an efficient solution for large-scale emotion recognition.
\end{itemize}

Our paper is organized as follows. Section 2 introduces literature and reviews related works. In Section 3, our proposed model is explained in detail. Experimental validations and result analysis are described in Section 4, along with the details of implementation. The slights and discussions are expressed in Section 5. Finally, we conclude this paper in Section 6 by summarizing the attributions and future prospects. Our codes have been released at https://github.com/guanghaoyin/RTCAN-1D.

\section{Related Work}
\subsection{Emotion Representation}
When we focus on emotional recognition, the emotions should be quantitatively represented. Psychologists model the emotion in two types: (1) dividing the emotion into discrete classes and (2) using the multidimensional emotion space.

The discrete emotion representation uses several keywords to describe emotion rates. Plutchik~\cite{plutchik1982a} proposed a wheel model that includes eight basic emotions. In his model, the stronger emotions are in the center, while the weaker emotions are surrounded. More complex emotions can be formed by a mix of basic ones. Then, Damasio~\cite{damasio1994descartes} categorised the emotions as primary (deriving from innate fast and responsive behaviour) or secondary (deriving from cognitive thoughts). Izard~\cite{izard2007basic} extended to 10 basic emotions. However, the discrete description is unsuitable for analyzing complex emotions, because some mixed emotions cannot be precisely expressed in words.

The second mainstream presented a certain degree of specific emotional level in the multidimensional space of emotion. A first attempt was proposed by Wundt~\cite{wundt1921vorlesungen}, where emotions were described in a pleasure-displeasure, excitement-inhibition, tension-relaxation 3D space. In 1979, Russell~\cite{russell1980circumplex} postulated a two-dimensional model, spanned by the axes valence
and arousal. The valence reflects the value of the positive or negative state, and the arousal measures the active or calm degree~\cite{dawson2007electrodermal}. Now, the V/A model becomes one of the most frequently employed model because it can be
easily assessed using the Self-Assessment Manikins (SAM). The three affective datasets used in this paper are all measured by V/A space.

\subsection{Multimodal Fusion for Emotion Recognition}
Multimodal information fusion technique is used to improve the performance of the system by integrating different information on the feature level, intermediate level, or decision level.  To improve the robustness and performance of emotion recognition, previous researchers fuse the features from different physiological signals~\cite{kim2008emotion,jerritta2011physiological}, including ECG, RSP, EMG, EDA, and EEG because the fused signals contain more implicit information with the different sampling rate and more channels. But collecting multi-signals requires many conductive electrodes placed along the subject's scalp. The user-unfriendly acquisition limits practical applications. Therefore, some works added external media messages, which have a great relationship with human emotion. Kim \textit{et al.}~\cite{kim2007bimodal} systematically present the robust emotion recognition fusing the speech and biosignals. Liu \textit{et la.}~\cite{liu2017multi} combined the evoked video and EEG signals to predict the subject's emotional state. Koelstra \textit{et la.}~\cite{koelstra2013fusion} combined the facial expression with EEG signals. And Thammasan \textit{et la.}~\cite{thammasan2017multimodal} fused the EEG and musical features. There is no effective exploration for the fusion of EDA and music signals, especially with large-scale subjects. Hence, our paper focuses on this issue.

\subsection{EDA Processing}
\label{eda decomposition}
The EDA signal is widely used in the psychology or physiology field. The approaches to process EDA signal include Bayesian Statistics~\cite{shukla2019feature}, dictionary learning~\cite{kelsey2018applications}, and decomposition~\cite{alexander2005separating}. We focus on the end-to-end general solution. Hence, complex processing is inappropriate. Therefore, we choose to decompose the EDA. The frequently used measurement of EDA is skin conductance (SC), which is composed of tonic, phasic components~\cite{boucsein2012electrodermal}. The tonic phenomena (t) reflect slow drifts of the baseline level and spontaneous fluctuations in SC. The phasic phenomena (r) is generally a short-time rapid fluctuation and reflects the response to external stimulation~\cite{boucsein2012electrodermal}. For realistic tasks, the measurement noise of equipment is also taken into consideration.

There have been several mathematical solutions associated with a stimulus for EDA decomposition in recent decades. Referring to Alexander $et\, al.$ \cite{alexander2005separating}, the well-known assumption is that the SC equals to the convolution between discrete bursting episodes of sudomotor nerve activity (SMNA) and biexponential impulse response function (IRF):
\begin{equation}
SC = SMNA * IRF.
\end{equation}

There exists some methods for EDA decomposition. Discrete deconvolution analysis (DDA)~\cite{benedek2010decomposition} claims the nonnegativity of the driver and maximal compactness of the impulses to decompose EDA through nonnegative deconvolution. Continuous decomposition analysis (CDA)~\cite{benedek2010continuous} improves DDA with the IRF called Bateman function to conduct continuous decomposition. Convex Optimization-Based EDA Method (CvxEDA) \cite{greco2015cvxeda} models the phasic, tonic, noise components and applies convex optimization to solve the maximum a posteriori (MAP) problem. CvxEDA provides a novel method to decompose EDA without preprocessing steps and heuristic solutions \cite{greco2016arousal}.

\subsection{Attention Mechanism}
In human proprioceptive systems, attention generally provides a guide to focus on the most informative components of an input \cite{hu2018squeeze}. Recently, attention models improve the performance of deep neural networks for various tasks, ranging from video classification \cite{wang2018non} to language translation \cite{vaswani2017attention}. Pei $et\, al.$ \cite{pei2017temporal} proposed a temporal attention module to detect salient parts of the sequence while ignoring irrelevant ones in sequence classification. Wang $et\, al.$ \cite{wang2018non} proposed a portable nonlocal module for spatial attention to capture long-range dependencies in video classification. Hu $et\, al.$ \cite{hu2018squeeze} proposed SENet to incorporate channel-wise relationships to achieve significant improvement for image classification.

To the best of the authors' knowledge, the attention mechanism has not been proposed in physiological-based emotion recognition. As the mixed 3-channel input is sequential, each channel may contribute differently to the final prediction in different temporal range. In this work, we use an efficient channel-wise and temporal attention mechanism based on 3-channel mixed EDA characteristics.

\section{Proposed Method}
\subsection{Network Framework}
The great success of CNN frameworks depends on 2D convolution in signal processing and image domain~\cite{he2016deep}. Our previous work~\cite{yin2019user} in EDA-based emotion recognition with the PMEmo dataset~\cite{zhang2018pmemo} also conducted 2D convolution. However, it is a detour to transform the 1D signal to 2D matrix just for fine-tuning pre-trained 2D CNN backbones. Moreover, if using the 2D convolution, the pre-precessing should be conducted to rearrange the 1D EDA sequences and transform them to 2D matrices as our previous work did~\cite{yin2019user}, which is time-consuming and increases the computations. Accordingly, we directly use the 1D residual CNN framework to process EDA signals in this work.

The overall architecture of the proposed method is illustrated in Fig. \ref{architecture}. Let the vector $X_T= \{x_1,x_2,...,x_t\}$ be the input EDA sequence with $t$ points. $ Y_{class} = \{y_1,...,y_c \}$ denotes the ground truth for $c$-classes task. If the emotion state belongs to the $i^{th}$ class, $y_i=1$ and $y_j=0$ for $j \neq i$. Our goal is to establish the relationships between input EDA and emotion state, specifically decreasing the distance between ground truth label $Y_{class}$ and predicted result $\hat{Y_{class}}$.

\begin{figure*}
\centering
\includegraphics[scale=0.3]{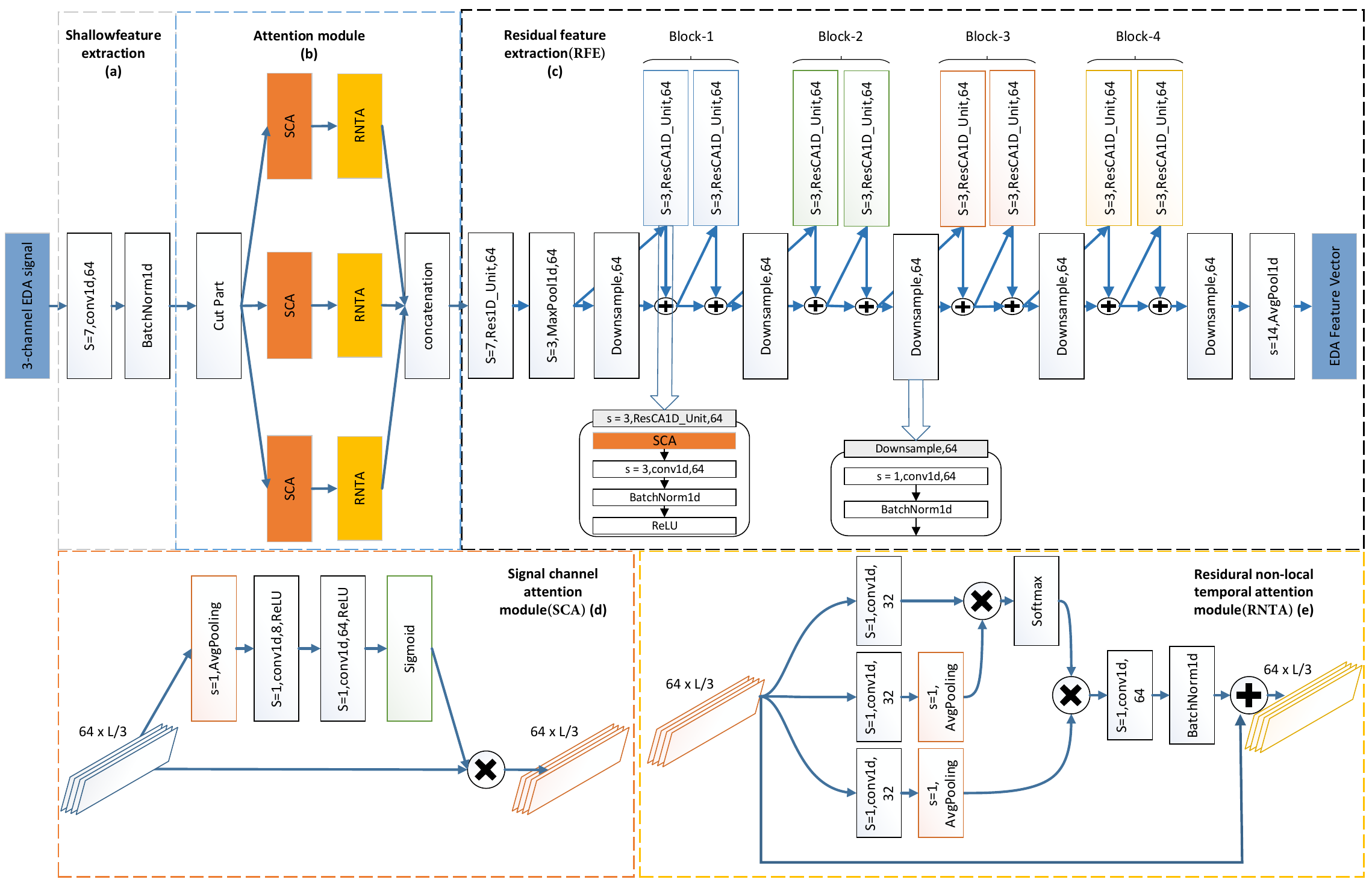}\\
\caption{The structure of the residual temporal and channel attention group (RTCAG) consists of (a) shallow feature extraction, (b) attention module, and (c) residual feature extraction (RFE). Specifically, the attention module contains (d) signal channel attention module (SCA) and (e) residual nonlocal temporal attention module (RNTA). It should be emphasized that the three blocks of The SCA and RTNA are the same one, respectively, which is recurrently used to reduce the modules' parameters.}
\label{rtcag}
\end{figure*}

First, after z-score normalization, we conduct CvxEDA decomposition. The normalized EDA signal (y) is composed of three N-long column vectors: the phasic (r) and tonic (t) signal plus the noise component ($\varepsilon$):
\begin{eqnarray}\label{equ:y}
	y = MA^{-1}p + B\lambda + Cd + \varepsilon,
\end{eqnarray}
where $r = MA^{-1}p$ and $t = B\lambda + Cd$. Then, with the transcendental knowledge of physiology and taking logarithmic transtorm, the MAP problem can be rewritten as a standard Quadratic-Programming (QP) convex form and can be solved efficiently using many available solvers (see more details in \cite{greco2015cvxeda}). For denoizing in the process of solving optimization problem, we discarded the prior probability of noise term. After CvxEDA decomposition, the 1-channel EDA signal is expanded to mixed 3-channel signal $X_{TC} = \{(x_{11},x_{12},x_{13}),...,(x_{t1},x_{t2},x_{t3}) \}$. As shown in Fig.\ref{rtcag}, in order to reduce the computational complexity, we uniformly clip shallow feature to 3 parts and the clipped signals are subsequently fed into the residual temporal and channel attention group (RTCAG) with shared attention weights. The EDA feature vector is extracted by three parts: shallow feature extraction, attention module, and residual feature extraction.

One convolutional layer extracts shallow feature $F_{SF}$ as
\begin{eqnarray}
  F_{SF}=H_{SF}(X_{TC}),
\end{eqnarray}
where $H_{SF}(\cdot)$ denotes convolution and batch-normalization operation. Then, the attention module reweights the shallow feature and focuses on more useful parts in temporal and channel-wise dimensions as
\begin{eqnarray}
  F_A=H_A(F_{SF}),
\end{eqnarray}
where $F_A$ is rearranged feature and $H_A(\cdot)$ is the attention module respectively. To mine discriminative representations, we design the residual feature extraction ($H_{RF}(\cdot)$) with ResNet-18 backbone~\cite{he2016deep} to output a EDA feature vector ($F_{EF}$):
\begin{eqnarray}
  F_{EF}=H_{RF}(F_A).
\end{eqnarray}
It should be emphasized that there exists other excellent CNN architectures such as VGG \cite{simonyan2014very}, or DenseNet \cite{huang2017densely} to mine deep features, but a comparison of different CNN frameworks is not the focus of our research. To be specific, for dozens of people involved EDA dataset, the subjects' specificity is not so complicated that the RTCAN-1D can handle the single EDA input, which means $F_{FF}=F_{EF}$. For the large-scale PMEmo dataset with hundreds of subjects, we fuse the EDA feature with an external static media feature ($F_{MF}$). Under this circumstances, the full feature vector is the concatenation of EDA feature and media feature:
\begin{eqnarray}
  F_{FF}=Cat(F_{EF},F_{MF}).
\end{eqnarray}

Finally, 	3 fully-connected layers followed with 3 ReLU functions, termed as ($H_{CF}(\cdot)$), are fed with the full feature vector ($F_{FF}$) and output the classifying vector ($\hat{Y_{class}}$) with the Softmax function
\begin{eqnarray}
  \hat{Y_{class}}=Softmax(H_{CF}(F_{FF})).
\end{eqnarray}

\subsection{Residual Temporal and Channel Attention Group}
Now we show our residual temporal and channel attention group (RTCAG) (seen in Fig. \ref{rtcag}). After the shallow feature extraction, the feature map is processed with the series structures of signal channel attention module (SCA), residual nonlocal temporal attention module (RNTA), and residual feature extraction (RFE) to output an EDA feature vector. The purpose of the attention mechanism is to adaptively reweight the shallow feature for more useful messages in the channel and temporal dimensions. Hence, the SCA and RNTA are plugged before RFE.

\subsubsection{Signal Channel Attention Module}
\label{SCA}
For multichannel input, different channels make different contributions to a certain target. We involve a channel-wise attention mechanism to exploit the interchannel relationship and adaptively reweights them. It recently has been a hotspot in various researches such as image classification \cite{hu2018squeeze}. By achieving high performance and robust to noisy inputs, the channel-wise attention mechanism is mobile and portable for different tasks. Referring to various channel-wise attention module, our SCA is adapted from the simplest structure of Squeeze-and-Excitation Block in \cite{hu2018squeeze} and replaces the convolution with 1D operation to fit signal inputs (see in Fig. \ref{rtcag}(d)). The multiplication operations of attention modules (seen in Fig. \ref{rtcag}(c)(d)) cause the unacceptable computational burden, especially with large feature maps. To improve the training efficiency, we consider to uniformly clip the long-temporal shallow feature to smaller groups in sequence dimension. The clipping method can reduce the length of each feature sequence, then reduce the computational complexity of multiplication for each forward propagation. In order not to involve more network parameters, we also didn't add the number of attention modules but use one module with shared parameters to recursively process the clipped parts. With sufficient GPU memory, this recursive process can also execute in parallel. The number of clipping parts is considerable because an inappropriate division leads to the large size for each clip, and the detailed division will cause the gradient and optimal issues for the SCA. Referring to \cite{dawson2007electrodermal}, the EDA signal evoked from stimulations passes through a procedure of the latency, rise time, half recovery time. Although these intervals are not homogeneous in time, the physiological knowledge guides us to roughly divide the shallow feature to 3 parts in sequence dimension: $F_{SF}=[F_{SF_1},F_{SF_2},F_{SF_3}]$, where $F_{SF} \in R^{C \times L}$ and $F_{SF_1},F_{SF_2},F_{SF_3} \in R^{C \times L/3}$.

We first aggregate the temporal information from the clipped feature with the average-pooling operation to get the temporal average-pooled features $F_{avg}^{L/3}$. Then the descriptor is forwarded to a multilayer perceptron (MLP) with only one hidden layer. To reduce parameter overhead, the channel of average-pooled features through the hidden layer $W_0$ is decreased by the reduction ratio $r$ to $R^{C/r \times L/3}$, then recovered by the latter layer $W_1$. The sigmoid activation function generates the normalized channel attention weight $W_R \in R^{C \times L/3}$ between 0 to 1. Finally, the original shallow feature is reweighted by the multiplication with the attention weight of the channel. Overall, the reweighted channel-wise feature is computed as
\begin{eqnarray}
  F_{CA_i} = Sigmoid(W_1(W_0(F_{avr}^{L/3}))) \ast F_{SF},
\end{eqnarray}
where $i=1,2,3$ and $ F_{CA}=[F_{CA_1},F_{CA_2},F_{CA_3}]$ is forwarded to the RNTA to mine temporal-range dependencies.

\subsubsection{Residual Nonlocal Temporal Attention Module}
Nonlocal operation is first proposed as a neighborhood filtering algorithm in the image domain~\cite{buades2005non}. It can involve the long-range position contribution to the filtered response of a certain position and control the contribution under the guidance of appearance similarity. At the advantage of capturing long-range relationships, the nonlocal neural network has been validated effective in video classification, object detection, instance segmentation, and keypoint detection~\cite{wang2018non}.
The great generality and portability motivates us to apply nonlocal operation in the recognition of psychological signals based emotions.

However, as described in Section~\ref{SCA}, the traditional attention module will multiply the computational complexity for large-size features. Therefore, the RNTA module performs a piecewise nonlocal operation. It takes the three parts of clipping features from the SCA as the inputs. Similar to the SCA, the RNTA is also trained recursively to reduce the model complexity.

Referring to \cite{buades2005non,wang2018non}, the generic nonlocal operation in deep neural networks can be defined as
\begin{eqnarray}
  \hat{x_i}=\frac{1}{c(x)}\sum{f(x_i,x_j)g(x_j)},
\label{18}
\end{eqnarray}
where $x$ represents the input feature map, $i$ is the target index of an output position in time sequence, and $j$ is the index of all possible locations which contribute to the filtered response $\hat{x_i}$. The output $\hat{x_i}$ has the same size as input $x$. Specifically, the function $g(\cdot)$ changes the representation of the input feature. The function $f(\cdot)$ computes the scale between $i$ and all $j$. The function $c(\cdot)$ is the normalization function before the output. The formulations of functions $c(\cdot)$, $g(\cdot)$ and $f(\cdot)$ depend on the specific network structure.

In our module (seen in Fig. \ref{rtcag}(e)), we set 1D convolution operation and an average-pooling operation with kernel of size 1 to conduct linear embedding as
\begin{eqnarray}
  g(x_j)=AvgPool(W_g x_j).
\end{eqnarray}
For pairwise function $f(\cdot)$, there exists various formulations to the relationship between long-range indexes like Embedded Gaussian, Concatenation, or, etc. Wang $et\, al.$ \cite{wang2018non} has validated that the performance of nonlocal operation is insensitive to the instantiations of function $f(\cdot)$ and $g(\cdot)$. Hence, we choose the commonly used Embedded Gaussian function~\cite{vaswani2017attention} and define $f(\cdot)$ as
\begin{eqnarray}
  f(x_i,x_j)=e^{\theta (x_i)^T \phi (x_j)},
\end{eqnarray}
where $\theta(\cdot)$ and $\phi(\cdot)$ are the convolutional layer. Adapted from \cite{wang2018non}, we set $c(x)=\sum_{\forall j}f(x_i,x_j)$, then $\frac{1}{c(x)}f(x_i,x_j)$ equals to the softmax operation. So, the nonlocal operation is computed as:
\begin{eqnarray}
  \hat{x}=softmax(x^T W^T_{\theta} AvgPool(W_{\phi}x)) AvgPool(W_g x).
\label{21}
\end{eqnarray}
Finally, we combine the nonlocal operation and the residual connection to get the output of attention module ($F_{A_i}$):
\begin{eqnarray}
  F_{A_i}= F_{RNTA_i}=W_wF_{NL_i} + F_{CA_i},
\end{eqnarray}
where $F_{CA_i}$ is the clipped input from the SCA module, $F_{NL_i}$ is calculated by Eq.(\ref{21}), $W_w$ denotes the 1D convolution operation followed with the batch normalization and $F_A=[F_{A_1},F_{A_2},F_{A_3}]$ is the reweighted output of our attention module.

\subsubsection{Residual Feature Extraction}\label{3.3.3}
The structure of EDA feature extraction module is shown in Fig.\ref{rtcag}(c). Our residual feature extraction (RFE) applies the ResNet-18 \cite{he2016deep} as the backbone. Considering the training efficiency, the dimension of signal sequences, and the complexity of physiological signals, which are less obvious to perceive the relationship with labels visually, we make two modifications: (1) replacing all the 2D convolutional layer with 1D convolutional operation for directly mining features from 1D signal; (2) simplifying the basic residual block that only conducts stacking 1D convolutional layer, batch normalization layer and ReLU nonlinear activation function for one time; (3) adding a signal channel attention layers at before the 1D convolution to mine the channel relationships between the deep features. Finally, the module extracts the EDA feature vector $F_{EF}$ from attention reweighted features $F_A$. Specifically, we found that in small-scale datasets, the SCAs of residual blocks will increase the fitting ability of RTCAG-1D and cause the overfitting drop-off (seen in Section \ref{4.5.5}).
Therefore, we removed the SCA in RTCAG for DEAP and AMIGOS datasets.

\subsubsection{Multimodal Fusion Classification}\label{mul}
Compared with image- or speech-recognition tasks, the complicated shapes of physiological curves show that physiological signal-based emotional recognition is a tough problem. It is not intuitive to perceive the relationship between the emotional state and the input curve. Moreover, personal specificity determines the intricate difference of nonemotional individual contexts between people \cite{kim2008emotion}, especially in datasets with large-scale subjects. Towards this end, we conduct multimodal fusion by adding some static benchmarks, which are accessible and have a great relationship with an emotional state. For the current multimodal emotion dataset, the media, such as music or video, is widely used to raise emotional states with affection involvement. The current largest PMEmo dataset used music that can be conveniently used as the prior benchmark. We simply applied the open-source toolkit openSMILE to extract
coarse music features because the fine EDA features will give us more information about subjective experience. As Fig. \ref{architecture} illustrates, the EDA feature vector and music feature vector are concatenated and forwarded to the multi-linear classification layers. The classifier consists of three linear layers, followed by a ReLU function. The output channels are 256, 128, and the class number, respectively. Finally, the output vector is normalized by the softmax function, and the maximum one is chosen as the predicted result.

\section{Dataset Analyses}
As the performance of the classifier is relevant to the data distribution. In this section, we will briefly introduce and analyze the characteristics of three affective datasets, including DEAP \cite{koelstra2011deap}, AMIGOS \cite{correa2018amigos}, and PMEmo \cite{zhang2018pmemo}, which can explain the experimental results in Section~\ref{multimodal analyses}. It should be emphasized that the music signals are only available on PMEmo. The DEAP and AMIGOS are mainly used to evaluate the RTCAG for EDA feature extraction, and we compare RTCAG with the existing single EDA-based classifier on these two datasets. The multimodel evaluation by fusing music and EDA is conducted on the PMEmo dataset.

\subsection{Datasets Introduction}

\textbf{DEAP.} The database for emotion analysis using physiological signals (DEAP) is one of the most explored datasets for affective computing. The 32 participants (19-37 years, 50\% females) separately watched 40 one-minute long videos. Meanwhile, the central nervous signal EEG and peripheral nervous signals including electromyogram (EMG), electrooculogram (EOG), skin temperature (TMP), galvanic skin response (GSR), blood volume pulse (BVP), and respiration (RSP) were recorded at a 512 Hz sampling rate and later down-sampled to 256Hz. Finally, the subjects performed the assessment of float subjective ratings on arousal, valence, liking, and dominance scales. From the dataset server, we downloaded ``data\_preprocessed\_python.zip", which has been downsampled to 128 Hz and segmented into 60-second trials. The 3-second pretrial baseline has been removed.

\textbf{AMIGOS.} The dataset for affect, personality, and mood research on individuals and groups (AMIGOS) consists of the short videos and long videos experiments: (1) the 16 clips of short videos were watched by 40 subjects; (2) the 37 subjects were separated to watch 4 long videos where 17 subjects performed in an individual setting and 5 groups of 4 people did in a group setting. During two experiments, physiological signals (EEG, ECG, GSR) were captured, where GSR signals are recorded at 128 HZ. Each participant rated each video in valence, arousal, dominance, familiarity, and linking and selected from 7 basic emotions. In our paper, we downloaded the pre-processed and segmented version from the ``data\_preprocessed\_matlab.zip."

\textbf{PMEmo.} The popular music dataset with emotional annotations (PMEmo) is our previous work~\cite{zhang2018pmemo}. To the best of our knowledge, it is the current largest emotion recognition dataset with EDA and music signal. The chorus excerpts clipped from 794 pop songs are collected as the emotion elicitation. After the relaxing procedure, the subjects listened to music, and the EDA signals were collected from subject’s finger at a 50-HZ sampling rate. Each chorus was listened by at least 10 subjects. Meanwhile, the subject conducted dynamic V/A discrete annotation and static V/A discrete annotation for each chorus. Discarding some bad cases, 7962 pieces of EDA signals from 457 subjects were finally collected. In this work, we use static annotations to represent the emotional state and fuse the EDA and the music signal for the recognition of large-scale emotions.

\begin{figure*}
\centering
  \subfloat[]{
  \includegraphics[width=3.3cm]{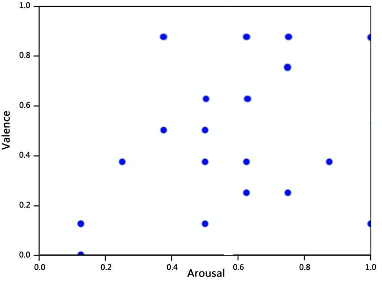}
  }%
  \subfloat[]{
  \includegraphics[width=3.3cm]{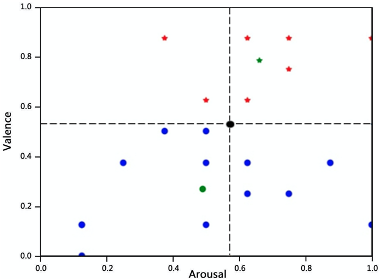}
  }%
  \subfloat[]{
  \includegraphics[width=3.3cm]{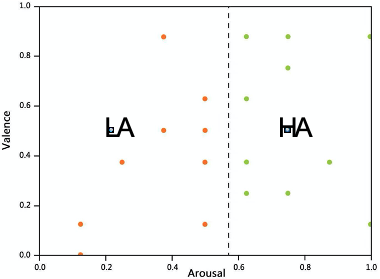}
  }%
  \subfloat[]{
  \includegraphics[width=3.3cm]{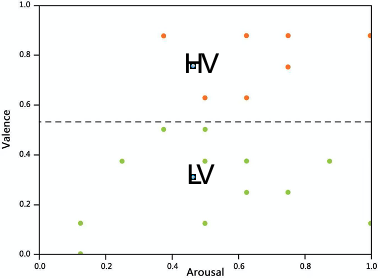}
  }%
\caption{Visualization of four-steps binary emotion label generation from one subject (ID:10031). (a) First, we collect annotations from a subject in the V/A emotional space; (b) then, the k-means clustering is applied to calculate two cluster centers and we also calculate their midpoint; (c) the high and low arousal states are separated by threshold; (d) finally, the high and low valence states are separated by threshold.}
\label{fig:threshold}
\end{figure*}

\begin{table}
\caption{The statistics of binary relabeled annotations generated by four-steps procedure in valence and arousal in three datasets.}
\label{statistics}
\begin{tabular}{cccc}
  \toprule
  \textbf{Dataset}&\textbf{V/A}&\textbf{Num of high class} &\textbf{Num of low class}\\
  \cline{2-4}
  PMEmo&Arousal&4520 (58.8\%)& 3170 (41.2\%)\\
  &Valence&4436 (57.7\%)& 3254 (42.3\%)\\
  \cline{2-4}
  DEAP&Arousal&680 (53.1\%)& 600 (46.9\%)\\
  &Valence&647 (50.5\%)& 633 (49.5\%)\\
  \cline{2-4}
  AMIGOS&Arousal&443 (56.2\%)& 345 (43.3\%)\\
  &Valence&394 (50.0\%)& 394 (50\%)\\
  \bottomrule
\end{tabular}
\end{table}

\subsection{Annotation Recreation}
The annotations from the above datasets are ranged from 1 to 9. As mentioned before, we attempt to conduct 2-class recognition in DEAP, AMIGOS, and PMEmo for a fair comparison with previous works. Therefore, the annotations should be relabeled to the appropriate classes.

Focusing on user-independent emotion recognition with large-scale subjects, we should pay more attention to personal specificity. Different emotion thresholds result in the same V/A score totally mapping to the opposite emotional state between different subjects. It should be noted that we involved the subject threshold just for generating proper subject-specific emotional classes. The user-independent model should not focus on a specific individual user but deal with greatly varying individual differences. But it does not mean that we can ignore the subject specificity when recreating the binary label from the original dispersed annotation. When mapping the 1-9 annotations to the binary classes, previous work~\cite{frantzidis2010classification} simply discretized the annotations into low ($\leq$5) and high ($\geq$5) V/A states. The reference~\cite{simonyan2014very} points out that since the subjective ratings also possess the nonstationarity and subject specificity, the fixed threshold may not be suitable for all individual preferences. If we attempt to validate the generalization of our work in large-scale data, personal threshold generating subject-specific emotional classes is essential. Specifically, inspired by the strategy in~\cite{frantzidis2010classification} for binary classification, we calculate the subject's emotion threshold with discrete annotations as the four-steps procedure. The example of one subject (Subject-ID: 10031 from PMEmo) is shown in Fig.~\ref{fig:threshold}. The statistics of each class after label recreation is shown in Table~\ref{statistics}. It validates the relabeled processing will not cause imbalanced label distribution.

\begin{table}
  \caption{The statistics of correlation between the original annotated arousal and valence scores in three datasets.}
  \label{correlation}
  \begin{tabular}{ccc}
  \toprule
  \textbf{Dataset}&\textbf{Correlation}&\textbf{P-Value}\\
  \cline{2-3}
  AMIGOS&$r=0.56$~\cite{shukla2019feature}&$p\leq.001$~\cite{shukla2019feature}\\
  DEAP&$r=0.15$&$p\leq.001$\\
  PMEmo&$r=0.52$&$p\leq.001$\\
  \bottomrule
  \end{tabular}
\end{table}

\subsection{Annotation Correlation}
\label{annotation correlation}
 Previous works point out the EDA signal is relevant for arousal classification~\cite{torres2013feature,shukla2019feature}. However, the results reported by~\cite{correa2018amigos,shukla2019feature,ganapathy2020convolutional} show that the classification performance for the valence and arousal dimensions does not diverge considerably.
Shukla~\textit{et al.}~\cite{shukla2019feature} reveals that the subject preferences of V/A annotation scores greatly influence the gap of V/A performance. In the specific dataset, the subject tends to annotate the valance score associated with arousal score, where the correlations of V/A scores are high. To systematically analyze the experiment results in Section~\ref{multimodal analyses}, we should first analyze the annotation correlations in DEAP, AMIGOS, and PMEmo datasets. Following~\cite{shukla2019feature}, we calculate the Spearman correlation coefficient $r$ between the valence and arousal annotations. The significant test of p-Value is also conducted. The statistics are listed in Table~\ref{correlation}. Compared with DEAP dataset ($r=0.15$), there are relatively higher correlations ($r=0.56/0.52$) between the annotated arousal and valence scores of AMIGOS and PMEmo with the strong p-value evidence ($p\leq .001$). It means the valance annotations in those two datasets have been influenced by arousal annotations. For DEAP dataset, the subject has no preference to annotate the valence scores associated with arousal annotations. The characteristics of annotation can explain that our RTCAG with single EDA delivers the accurate valence recognition with AMIGOS and PMEmo datasets, and the performance of valence variable is much lower than arousal one in DEAP dataset. For the multimodel fusion of RTCAN-1D, the involvement of music overcomes the defect of EDA signals to achieve high performance in both the valence and the arousal dimensions.

\section{Experiments}
In this section we introduce the training settings briefly. Then, systemic multimodal analyses are conducted. Because the music and EDA signals from large-scale subjects are only available from PMEmo, the analyses with single music and multimodal inputs are conducted on PMEmo. The DEAP~\cite{koelstra2011deap} and AMIGOS~\cite{correa2018amigos} with fewer subjects are used for single EDA-based emotion recognition. Because of the individual specificity of large-scale subjects, we also conduct the incremental subject experiments with single EDA inputs on PMEmo to observe the overfitting of CNN models. Moreover, we carry out the ablation study to present the effects of different components of RTCAN-1D. The comparisons with previous solutions prove that our proposed multimodal framework can efficiently solve those problems. More details are presented as follows.

\subsection{Training Settings}
For fair comparisons with other existing methods, the weights of the whole RTCAN-1D in both DEAP, AMIGOS and PMEmo datasets were initialized from the Gaussian distribution $N(0,0.01)$. We trained our model with cross-entropy loss and empirically selected the size of mini-batches as 256. The Stochastic Gradient Descent (SGD) was applied as the optimizer. Moreover, we set the initial learning rate to 0.001, which was decreased by 0.9 at every 15 epoches.

The model suggested was assessed with 10-fold cross-validation. The dataset was divided into 10 disjoint subsets with subject-independent scenarios. At each fold, one subset was set to the validation data for fine-tuning hyperparameters, one subset was set to the test set, and the rest were used for training. Repeated 10 times, all the subsets were tested, and the average metrics of all the 10 folds were calculated as the results. The performance of binary classification is evaluated with average accuracy and F1-score (the harmonic mean of precision and recall).

Before fed into a deep neural network, the EDA and music signals will be pre-processed. Processed by z-score normalization and CvxEDA, the EDA signal was split into tonic, phasic components. The signal denoizing could be accomplished by convex optimization. Then, the z-score normalization was also conducted for three EDA and music signals. Moreover, we just used EDA signals after 3-second because annotators need some preliminary to evoke their emotion~\cite{aljanaki2017developing}. For input size alignment, linear interpolation was also conducted to rescale EDA inputs.

\begin{table}
  \renewcommand\tabcolsep{1pt}
\caption{The multimodal analyses with different models, input signals, and different scale subjects in DEAP, AMIGOS, and PMEmo datasets.}
\label{dif_fea}
\begin{tabular}{cccccccc}
\toprule

\multirow{2}{*}{\textbf{Method}}&\multirow{2}{*}{\textbf{Dataset}}&\multirow{2}{*}{\textbf{Subject}}&\multirow{2}{*}{\textbf{Input signal}}&\multicolumn{2}{c}{\textbf{Valence}}&\multicolumn{2}{c}{\textbf{Arousal}}\\
&&&&\textbf{Ave-acc}&\textbf{F1-Score}&\textbf{Ave-acc}&\textbf{F1-Score}\\

\midrule
RTCAG&DEAP&32&EDA&77.15\%&79.18\%&83.42\%&83.77\%\\
RTCAG&AMIGOS&40&EDA&75.87\%&74.48\%&78.20\%&77.56\%\\
\midrule
SVM&PMEmo-10\%&46&EDA hand-crafted feature&60.48\%&61.51\%&61.73\%&61.95\%\\
SVM&PMEmo-40\%&128&EDA hand-crafted feature&60.27\%&61.04\%&61.49\%&61.77\%\\
SVM&PMEmo-70\%&319&EDA hand-crafted feature&60.03\%&60.87\%&61.12\%&61.35\%\\
SVM&PMEmo&457&EDA hand-crafted feature&59.78\%&60.36\%&60.95\%&61.14\%\\
\midrule
3 layers CNN&PMEmo-10\%&46&EDA hand-crafted feature&65.68\%&65.33\%&65.74\%&65.90\%\\
3 layers CNN&PMEmo-40\%&128&EDA hand-crafted feature&62.85\%&63.26\%&63.37\%&63.40\%\\
3 layers CNN&PMEmo-70\%&319&EDA hand-crafted feature&62.39\%&62.92\%&62.55\%&62.88\%\\
3 layers CNN&PMEmo&457&EDA hand-crafted feature&61.64\%&62.30\%&61.98\%&62.14\%\\
\midrule
RTCAG&PMEmo-10\%&46&EDA&72.12\%&71.33\%&74.52\%&71.53\%\\
RTCAG&PMEmo-40\%&128&EDA&66.24\%&65.80\%&67.05\%&66.70\%\\
RTCAG&PMEmo-70\%&319&EDA&65.82\%&64.77\%&65.79\%&66.23\%\\
RTCAG&PMEmo&457&EDA&64.52\%&63.31\%&65.19\%&65.84\%\\
\midrule
SVM~\cite{yin2019user}&PMEmo&457&Music&70.43\%&75.32\%&71.49\%&76.36\%\\
\midrule
\textbf{RTCAN-1D}&\textbf{PMEmo}&\textbf{457}&\textbf{EDA + Music}&\textbf{79.68\%}&\textbf{82.45\%}&\textbf{83.76\%}&\textbf{86.12\%}\\

\bottomrule
\end{tabular}
\end{table}

\subsection{Multimodal Analyses}
\label{multimodal analyses}

\subsubsection{Single EDA Recognition}
\label{single EDA}

\begin{figure}
\centering
  \subfloat[Arousal ave-accuracies]{
  \includegraphics[width=3.4cm]{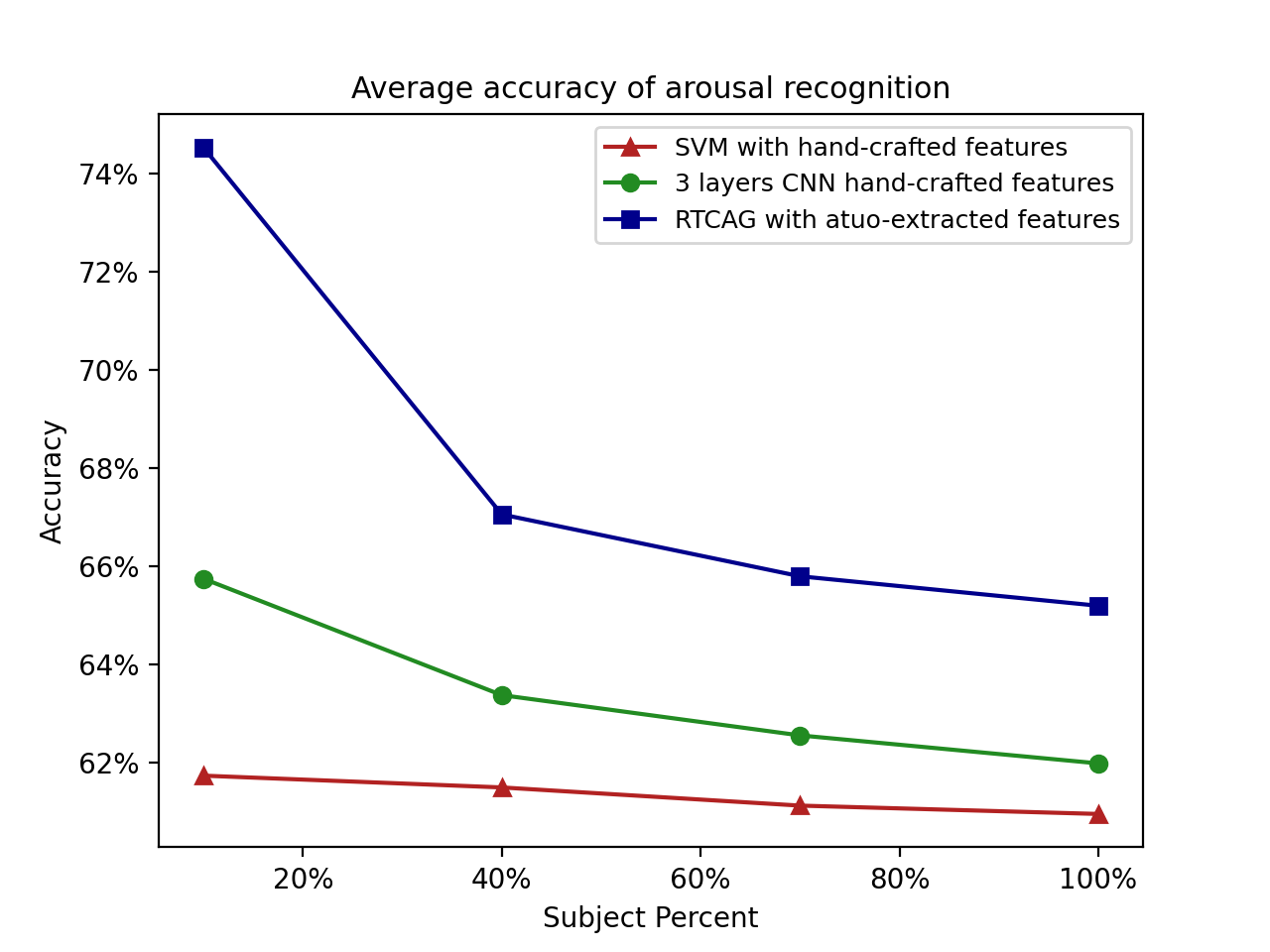}
  }%
  \subfloat[Arousal F1-scores]{
  \includegraphics[width=3.4cm]{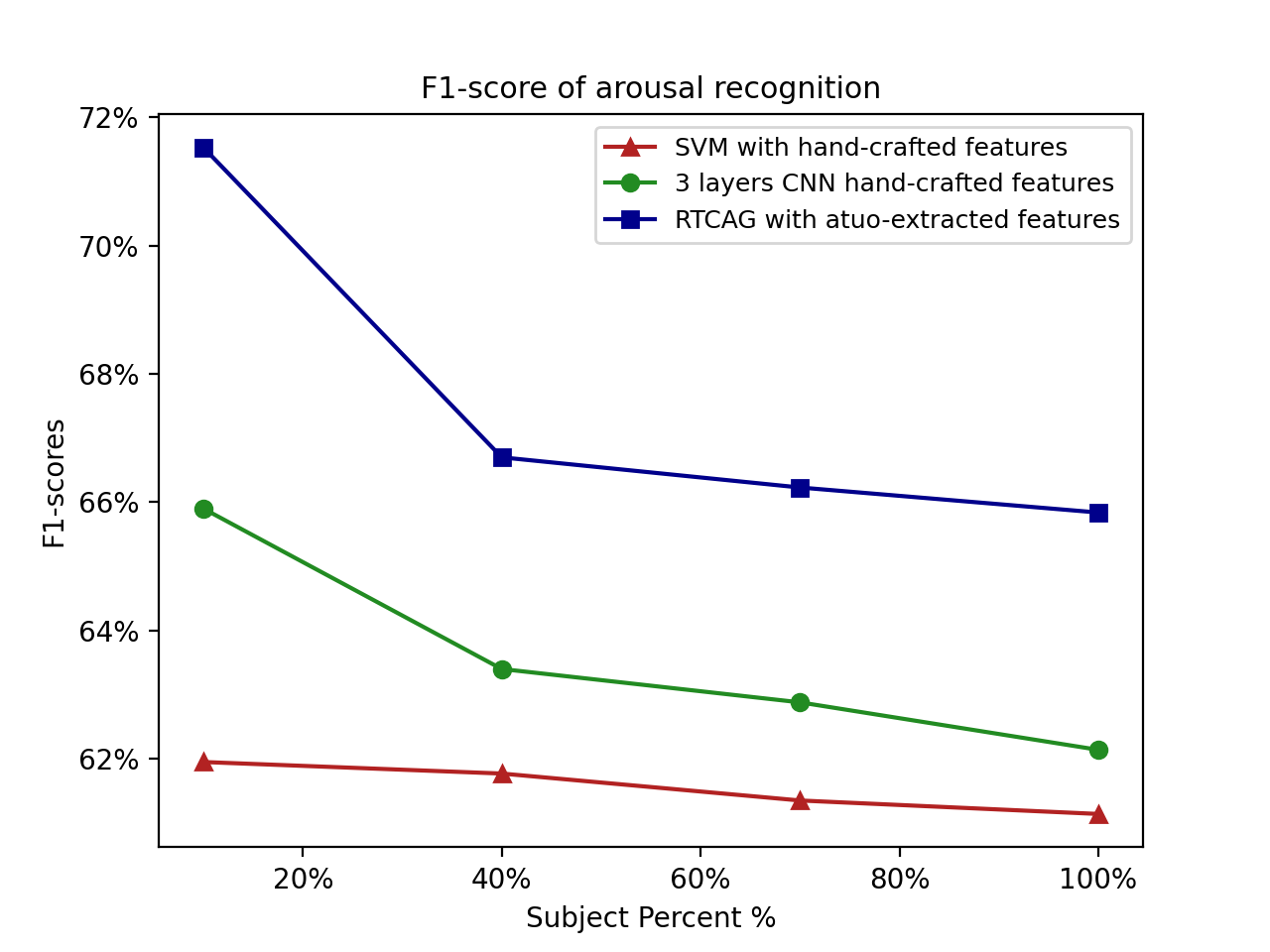}
  }%
  \subfloat[Valence ave-accuracies]{
  \includegraphics[width=3.4cm]{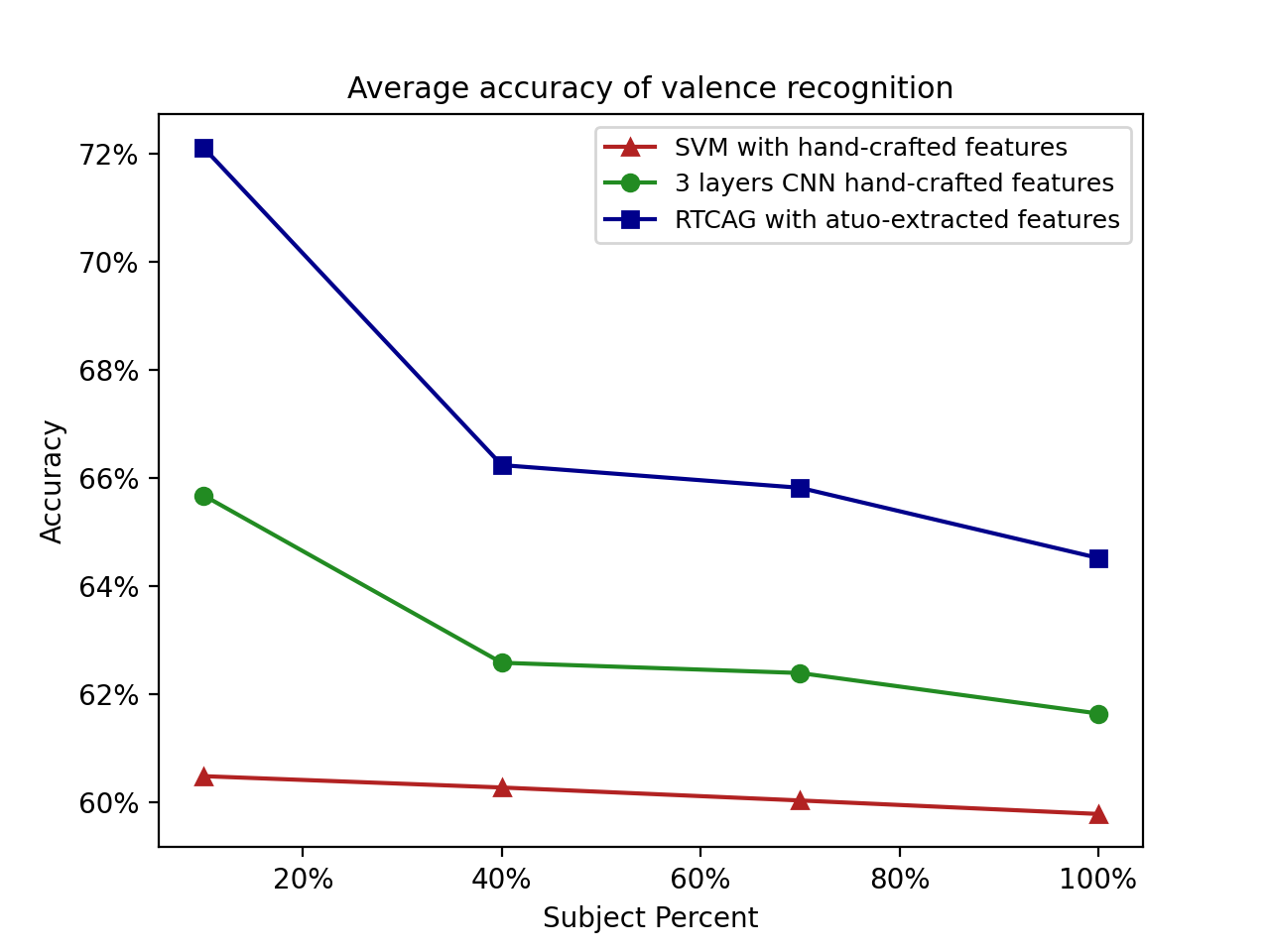}
  }%
  \subfloat[Valence F1-scores]{
  \includegraphics[width=3.4cm]{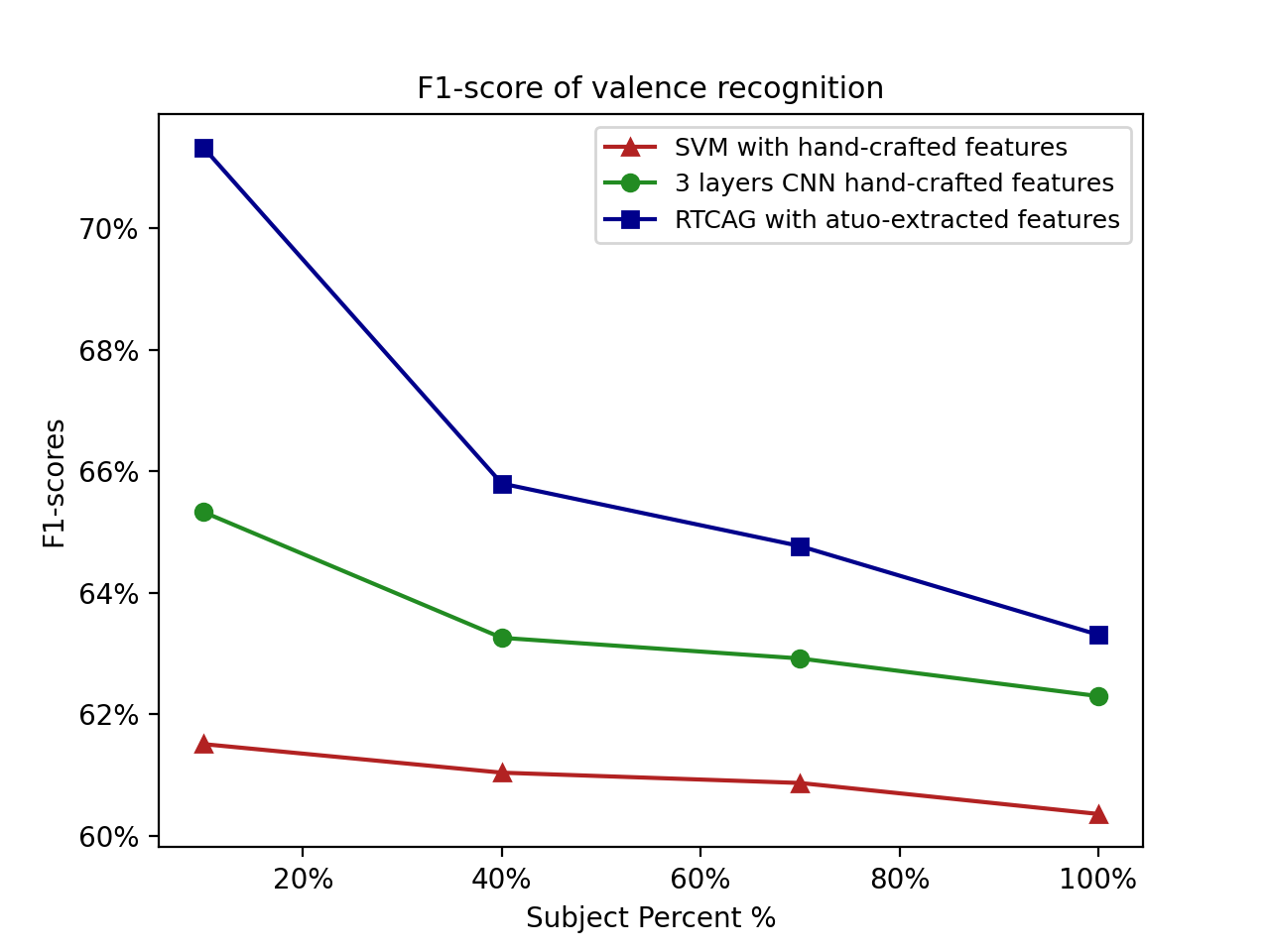}
  }%
\caption{The recognition results with subject incremental settings for arousal and valence dimensions in PMEmo dataset. The subjects are gradually increased from 10\% to 40\%, 70\%, and 100\%. It's clear that there is an significant drop-off from 10\%to 40\% subject percent and the decrease becomes gentle from 40\% to 100\%.}
\label{fig:subject_increase}
\end{figure}

To prove the efficacy of the proposed multimodal structure, we should first prove that useful EDA characteristics can be extracted by our RTCAG. The single EDA recognition is conducted in three datasets, including DEAP, AMIGOS, and PMEmo. After being pre-processed, the composed 3-channel signals were fed into RTCAG, and the series liner layers classified the EDA feature vectors.

The results of our RTCAG in DEAP, AMIGOS, and PMEmo are listed on the 1st, 2nd, and 14th rows of Table~\ref{dif_fea}. It is clear that our model has accomplished accurate recognition in DEAP and AMIGOS, which proves the EDA feature extraction of RTCAG is excellent. Moreover, because of the annotation correlations, the results for V/A dimensions in AMIGOS and PMEmo are relatively close, and the performance in DEAP diverges considerably, which is aligned with the dataset characteristics in Section~\ref{annotation correlation}. However, the performance in the large-scale PMEmo decreases 10\% with the increase in the number of participants, which intrigued us. Previous works of neural networks~\cite{lawrence1997lessons,srivastava2014dropout,lawrence2000overfitting} reveal the DNN models, especially deeper models with large parameters, may suffer overfitting when the inputs are complicated from various different domains. Considering the various subject specificity, we assume the RTCAG may suffer overfitting.

To validate our assumption, we conduct an incremental subject experiment. The subjects from PMEmo are gradually increased from 10\% to 40\%, 70\%, and 100\%. The benefit of hand-crafted features is that they have been shown to be stable by a variety of studies~\cite{shukla2019feature}. To systematically evaluate our model we add the traditional SVM classifier with time domain, frequency domain, and time-frequency domain features hand-crafted The 128 statistical features are calculated with the methods of~\cite{shukla2019feature}. And we refer to~\cite{correa2018amigos} to set the regularization parameter $C$ of the linear SVM as 0.25. Because the RTCAG is well-designed for 3-channel EDA signals, we wonder if the simple CNN with EDA feature engineering may suffer less over-fitting. Hence, we add another model in which the 128 statistical features are fed into a 3 layers CNN. The basic convolution layer of simple CNN keeps the same with RTCAG. Keeping the same training settings, the results of three models are shown on the 3rd row to 14th row of Table~\ref{dif_fea} and illustrated in Fig~\ref{fig:subject_increase}.

According to the results with single EDA input, we can make some analyses: (1) for RTCAG, when the number of subjects (PMEmo-10\%) is small, the recognition accuracies of RTCAG are competitive to DEAP and AMIGOS. When the involved subjects increase to hundreds, the recognition results decrease drastically. More individual specificity cannot help the model to converge but cause the RTCAG to overfit. (2) For 3 layers CNN with hand-crafted features, the model also suffers overfitting, but the decrease is not as sharp as RTCAG. It reveals that the EDA signals from hundreds of subjects are indeed complicated to cause overfitting even processed with feature engineering. On the other hand, the performance of 3 layers CNN is poorer than RTCAG, which means simpler architecture and fewer parameters limit the performance of CNN model. (3) As shown in Fig.\ref{fig:subject_increase}, the advantage of traditional SVM classifier with hand-crafted features is stable. When increasing the subjects, the SVM only shows a slight performance drop. However, it suffers underfitting and produces the worst recognition accuracy limited by the poor representation ability. (4) Although there is a slightly better performance for arousal, the recognition performance for the two variables in four subject scales does not diverge considerably, which is aligned with the high annotation correlation of PMEmo dataset as we have described in Section~\ref{annotation correlation}. According to the analyses, the defect of the single EDA input motivates us to include additional information through multimodal fusion.

\subsubsection{Single Music Recognition}
\label{single music}
This situation is common for user-independent emotion recognition: the same song listened by different people will stimulate the different emotional states, then get the opposite labels. The parameter optimization of deep learning model is strongly dependent on the loss between the predicted output and ground-truth label for backpropagation. When the same input data maps the total different ground-truth labels, the network optimization becomes very hard. If there exists many such data, the network training may even get corrupted. Therefore, for single music recognition, we directly used the result (SVM + Music) in our previous work~\cite{yin2019user}. The results are presented in the 7th row of Table~\ref{dif_fea}. We can make some analyses: (1) compared with single EDA input, the performance with single music input get improved for valence recognition because the EDA is less relevant to the valance state~\cite{torres2013feature,shukla2019feature} but the music has a strong relationship with subject’s valence/arousal moods~\cite{lin2011exploiting}. (2) For single music features, if considering subject specificity, the same music might lead to a completely opposite emotional state, which means the same music input maps to label 0 and 1 with different subject. (3) The single music input can reveal people's general emotion states at a certain extent. But it is not enough for user-independent emotion recognition. Although the SVM classifier with single music feature has achieved 70\% accuracies, it has limited ability to recognize the cases which map to different emotion labels because of individual differences. Towards this, we add the music feature as the static emotion benchmark to help the convergence of our RTCAN-1D.

\subsubsection{Fusing Music and EDA Recognition}
\label{multimodal}
As shown in the last row of Table \ref{dif_fea}, when fusing the music and EDA features, the proposed RTCAN-1D achieves the best performance. For single music recognition, the classifier could not recognize the individual specificity because the music feature had no relationships with a specific human. For single EDA recognition, the traditional classifier with feature engineering suffers underfitting, and our well-designed RTCAG model will get overfitting when large-scale subjects involve. The significant improvement of RTCAN-1D validates that the fusion of external static features from music and individual EDA features is a solution for large-scale emotional recognition.

\subsection{Ablation Analyses}
 To avoid the less-effective experiment repetition, we carry out most of the ablation studies in PMEmo. For some experiments of fitting ability, since the network structure has been slightly changed for different datasets, we present the results in DEAP, AMIGOS and PMEmo datasets.

\subsubsection{Res-block Configuration}
As Fig. \ref{architecture} and Fig. \ref{rtcag} illustrate, the residual feature extraction (RFE) is the principal part of the whole RTCAN-1D while the attention module, shallow feature extraction, and linear classifier are designed with the simplest structures. Therefore, the structure of RFE is the major determinants of computing cost. As explained in Section \ref{3.3.3}, we apply the Resnet-18 as the backbone and replace all 2D convolution operations with 1D version. Keeping the basic backbone of Resnet, the key problem addresses the depth of res-block (the stacking 1D convolutional layer, batch normalization, and ReLU) and the number of internal channels. The original settings of ResNet-18 are $d=2$ and $c=[64,128,256,512]$. We design res-block with different configurations. The trade-offs between each structure are shown in Table \ref{depth}.

In practice, we find that the best performance is given by the simplest structure. By using a deeper convolutional layer and more internal channels, the performance of RTCAN-1D get drop-off, and the network can be easily overfitting, which is more obvious in smaller scale DEAP and AMIGOS datasets. In general, a deeper network leads to better representation performance at the expense of space and time cost. However, when the inner class features are not consistent, and the number of training data scale is not large enough, such as millions of images for image recognition \cite{hu2018squeeze}, the complicated structure is harmful for EDA feature extraction. It is much clear in Table~\ref{depth}. Therefore, we applied the simplest structure $d = 1$ and $c=[64,64,64,64]$ for our res-block.

\begin{table}[h]
\caption{Trade-off between performance and the depth at each level of the residual block in PMEmo dataset.}
\label{depth}
\centering
\begin{tabular}{ccccc}
\toprule
\textbf{V/A}&\textbf{Depth}&\textbf{Channel}&\textbf{Ave-Acc}&\textbf{F1-Score}\\
\midrule
V&1&[64,64,64,64]&\textbf{79.68\%}&\textbf{82.45\%}\\
&1&[64,128,256,512]&75.42\%&78.80\%\\
&2&[64,64,64,64]&73.45\%&76.92\%\\
&2&[64,128,256,512]&72.84\%&75.47\%\\
\midrule
A&1&[64,64,64,64]&\textbf{83.76\%}&\textbf{86.12\%}\\
&1&[64,128,256,512]&79.31\%&80.83\%\\
&2&[64,64,64,64]&74.80\%&75.55\%\\
&2&[64,128,256,512]&73.43\%&74.71\%\\
\bottomrule
\end{tabular}
\end{table}

\begin{table}[t]
  \renewcommand\tabcolsep{1pt}
\caption{Comparison of different attention modules in DEAP, AMIGOS and PMEmo datasets.}
\label{mod}
\centering
\begin{tabular}{ccccccccc}
\toprule
&&\textbf{Basic} & \bm{$R_a$} & \bm{$R_b$} & \bm{$R_c$}& \bm{$R_d$}& \bm{$R_e$}&  \bm{$R_f$}\\
\midrule
\textbf{Module}&\textbf{Base}& \bm{$\surd$}&\bm{$\surd$}&\bm{$\surd$}&\bm{$\surd$}&\bm{$\surd$}&\bm{$\surd$}&\bm{$\surd$}\\
&\textbf{SCA\_Res}&&\bm{$\surd$}&&\bm{$\surd$}&&&\bm{$\surd$}\\
&\textbf{SCA}&&&\bm{$\surd$}&\bm{$\surd$}&&\bm{$\surd$}&\bm{$\surd$}\\
&\textbf{RTNA}&&&&&\bm{$\surd$}&\bm{$\surd$}&\bm{$\surd$}\\
\midrule
\multirow{4}{*}{\textbf{DEAP}}&\multirow{2}{*}{\textbf{Ave-Acc}}&V:76.28\%&V:74.45\%&V:76.61\%&V:74.02\%&V:76.53\%&\textbf{V:77.15\%}&V:73.33\%\\
&&A:81.94\%&A:78.48\%&A:82.45\%&A:78.04\%&A:82.71\%&\textbf{A:83.42\%}&A:77.51\%\\
&\multirow{2}{*}{\textbf{F1-score}}&V:78.21\%&V:75.40\%&V:77.98\%&V:74.97\%&V:78.33\%&\textbf{V:79.18\%}&V:74.60\%\\
&&A:81.95\%&A:77.87\%&A:82.40\%&A:77.33\%&A:82.80\%&\textbf{A:83.77\%}&A:76.95\%\\
\multirow{4}{*}{\textbf{AMIGOS}}&\multirow{2}{*}{\textbf{Ave-Acc}}&V:73.48\%&V:71.04\%&V:73.94\%&V:70.87\%&V:74.28\%&\textbf{V:75.87\%}&V:70.63\%\\
&&A:76.87\%&A:71.31\%&A:77.46\%&A:70.84\%&A:77.91\%&\textbf{A:78.20\%}&A:70.27\%\\
&\multirow{2}{*}{\textbf{F1-score}}&V:73.30\%&V:71.73\%&V:73.77\%&V:71.25\%&V:73.45\%&\textbf{V:74.48\%}&V:70.88\%\\
&&A:75.81\%&A:74.47\%&A:76.71\%&A:74.00\%&A:76.50\%&\textbf{A:77.56\%}&A:73.59\%\\
\midrule
\multirow{4}{*}{\textbf{PMEmo}}&\multirow{2}{*}{\textbf{Ave-Acc}}&V:74.82\%&V:75.93\%&V:75.67\%&V:77.40\%&V:75.61\%&V:76.26\%&\textbf{V:79.68\%}\\
&&A:77.34\%&A:81.16\%&A:80.43\%&A:82.05\%&A:80.73\%&A:80.92\%&\textbf{A:83.76\%}\\
&\multirow{2}{*}{\textbf{F1-score}}&V:78.50\%&V:80.22\%&V:79.15\%&V:81.60\%&V:79.33\%&V:80.75\%&\textbf{V:82.45\%}\\
&&A:82.53\%&A:85.20\%&A:84.13\%&A:85.62\%&A:84.66\%&A:85.15\%&\textbf{A:86.12\%}\\
\bottomrule
\end{tabular}
\end{table}

\subsubsection{Effects of Attention Module}
\label{4.5.5}
The proposed RTCAN-1D consists of basic extraction model features and attention modules. As explained before, simply designing a deeper residual feature extraction network cannot improve the representation ability. Therefore, we involve the attention mechanism that has achieved great success and proven to be portable in various tasks. There are two main components of our attention module: (1) signal channel attention module (SCA) mines the inter-channel relationship and rearranges the importance between different channels of shallow features, and (2) residual nonlocal temporal attention module (RNTA) contributes the long-range position to the filtered features and reweight the input by capturing the relationships between long-time signal clips. It can be seen that the same structures of SCA modules are used in two different positions of our model: one is for shallow feature mining in Fig. \ref{rtcag}(b) (noted as SCA) and the others are in res-blocks for residual feature extraction in Fig. \ref{rtcag}(c) (noted as SCA\_Res). Therefore, we should separately analyze those two SCA modules. To validate the effectiveness of the combining method of SCA, SCA\_Res and RNTA, we compared 6 variants of the model: the basic model without any attention module, the basic model respectively with single SCA and RTNA module for shallow feature mining, the basic model with two types of SCA modules, the basic model with two attention modules for shallow feature mining and the full model, which consists of the basic model and SCA, SCA\_Res and RNTA modules.

Experimental results with different attention modules in three datasets are shown in Table \ref{mod}. As introduced in Section \ref{single EDA}, the increase of subjects involved may increase the fitting ability of the model. In order to explore the appropriate structure of our network, we should analyze the results in the large-scale PMEmo dataset and two smaller scale DEAP and AMIGOS datasets, respectively.

In the large-scale PMEmo dataset, $Base$ only contains the ResNet backbone and multi-linear layers to classify the fused EDA+music feature vector, hence it gets the poor accuracy in both three datasets. The higher recognition accuracies from $R_a$ and $R_b$ prove the effectiveness of the individual attention subnet, compared with $Base$ model. When combining all the two modules, we can observe that the joint use of channel and temporal relationships $R_c$ produces further improvement. Reviewing the results in Table \ref{mod}, it should be emphasized that both SCA and RTNA are indispensable because all the sequential methods outperform the basic backbone when adding a single attention submodule in the large-scale PMEmo dataset. In smaller scale DEAP and AMIGOS datasets, it can be seen that the simplest $Base$ is not the worst model. When adding the SCA\_Res modules, three configurations $R_b$, $R_c$, $R_f$ get performance drop, the accuracies and F1-scores of which are even lower than $Base$. The most complicated structure $R_f$ produces the worst results. It means when we add SCA modules in each residual block, the fitting ability of RTCAG-1D gets increased and causes the overfitting drop-off. When we only add the SCA and RTNA to process the extracted shallow feature, the $R_e$ model achieves the best results. Therefore, we remove the SCA modules of residual blocks to avoid over-fitting in small-scale datasets. The experiments in this subsection also reveal that when we design the DNN structure for user-independent emotion recognition, the scale of the subject involvement should be carefully considered.

\subsubsection{Attention Visualization}
Recently, researchers in neural network interpretation have proposed some methods for qualitative analysis. One of the most popular methods is visualizing the weight distribution in the regions of input to show which parts are more important for predictions. Gradient-weighted Class Activation Mapping (Grad-CAM) \cite{selvaraju2017grad} is an excellent visualization method that applies back gradient computing to reversely acquire the important weight of the input positions in convolutional layers. In the image domain, the output mask of Grad-CAM is overlaid on the semitransparent input image in the form of a heatmap with a warm-to-cool color spectrum. As warmer color is applied, the highlighted pixels contribute more to the final prediction, and the model will focus more on these messages. For another model, if the Grad-CAM mask covers more target regions in the same input, the network's receptive field of the is larger. And generally, it means the network has superior representation ability.

To show the ability of our attention submodules intuitively, we compare the visualization attention weights of the 3 trained model in the preceding experiment of Section \ref{4.5.5}: $R_a$ (channel attention model), $R_b$ (temporal attention model) and $R_c$ (integrated channel-temporal attention model). In order to fit the proposed framework, we adapt Grad-CAM and start the guided backpropagation from the output layer of each attention submodule in order to invert 1D attention weights in the 3-channel EDA input.

\begin{figure}
\centering
  \subfloat[Channel attention]{
  \includegraphics[width=4.6cm]{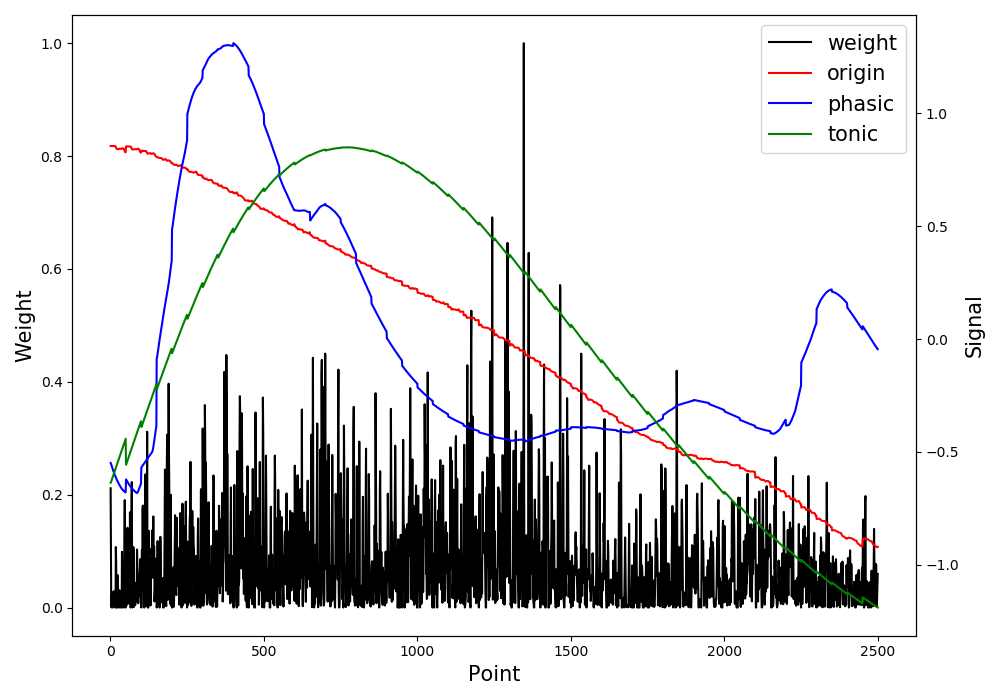}
  }%
  \subfloat[Temporal attention]{
  \includegraphics[width=4.6cm]{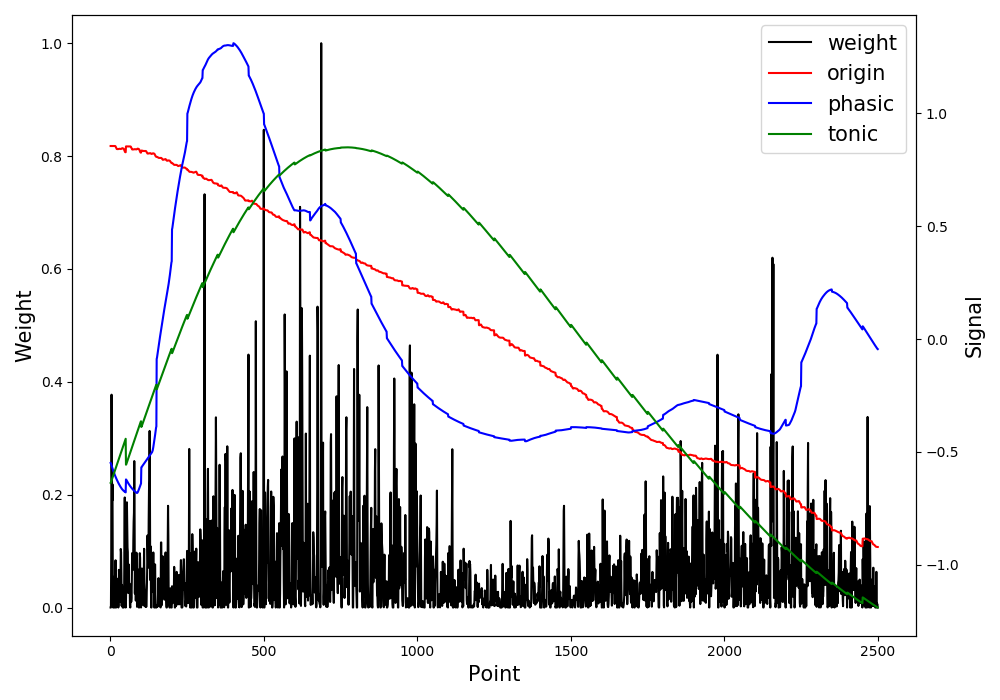}
  }%
  \subfloat[Channel-temporal attention]{
  \includegraphics[width=4.6cm]{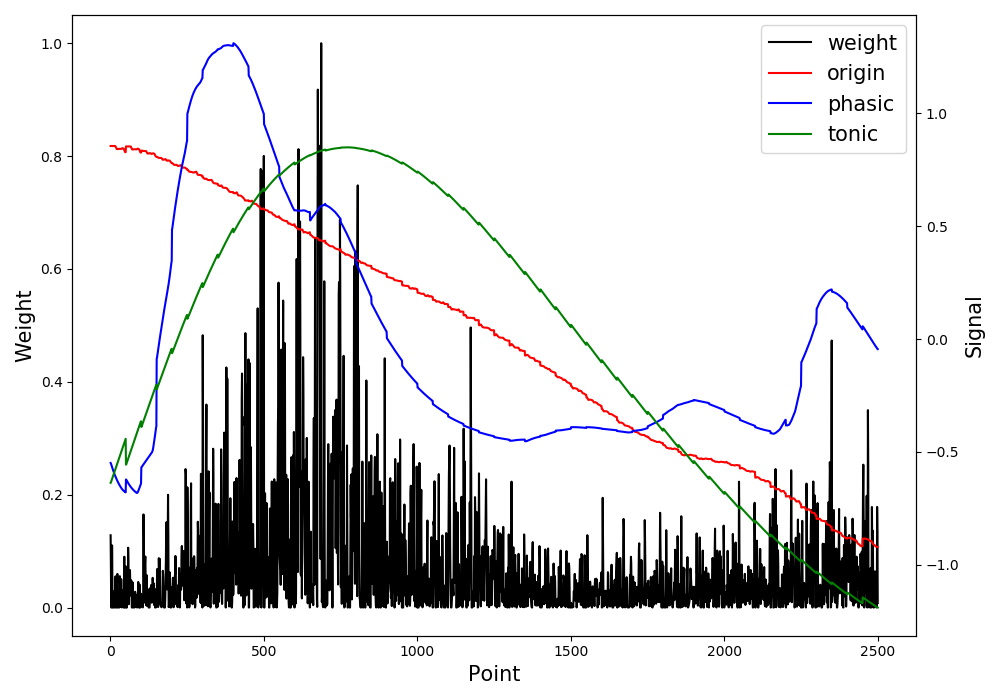}
  }%
\caption{The visualization of the curves of three signals and the column of attention weight in valence recognition with the random subject (Subject-ID: 100184, Music-ID: 1, fold: 10). The red, blue, and green curves note the origin, phasic and tonic signals, respectively. The black columns represent the inverse attention weight mapping from the output features of attention submodules to 3-channel input.}
\label{vc}
\end{figure}

\begin{figure}
\centering
  \subfloat[Channel attention]{
  \includegraphics[width=10cm]{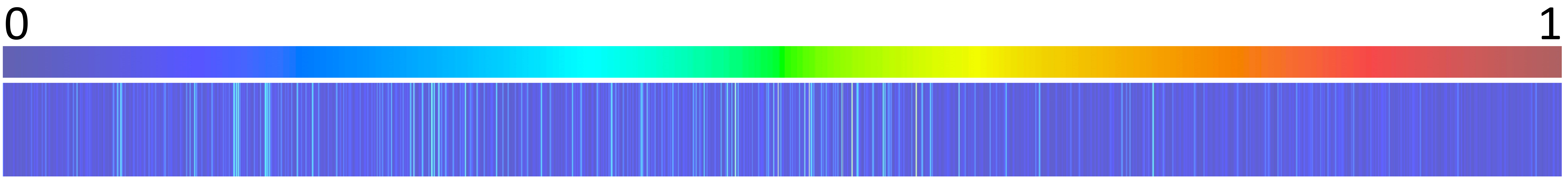}
  }%

    \subfloat[Temporal attention]{
  \includegraphics[width=10cm]{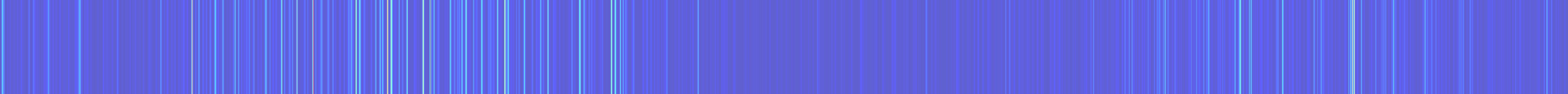}
  }%

    \subfloat[Channel-temporal attention]{
  \includegraphics[width=10cm]{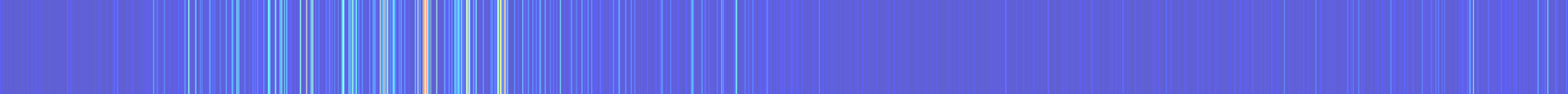}
  }%
\caption{The valence spectrogram visualization. The spectrograms in (a), (b), (c) separately correspond to the attention weights in Fig. \ref{vc}.}
\label{vh}
\end{figure}

\begin{figure}
\centering
  \subfloat[Channel attention]{
  \includegraphics[width=4.6cm]{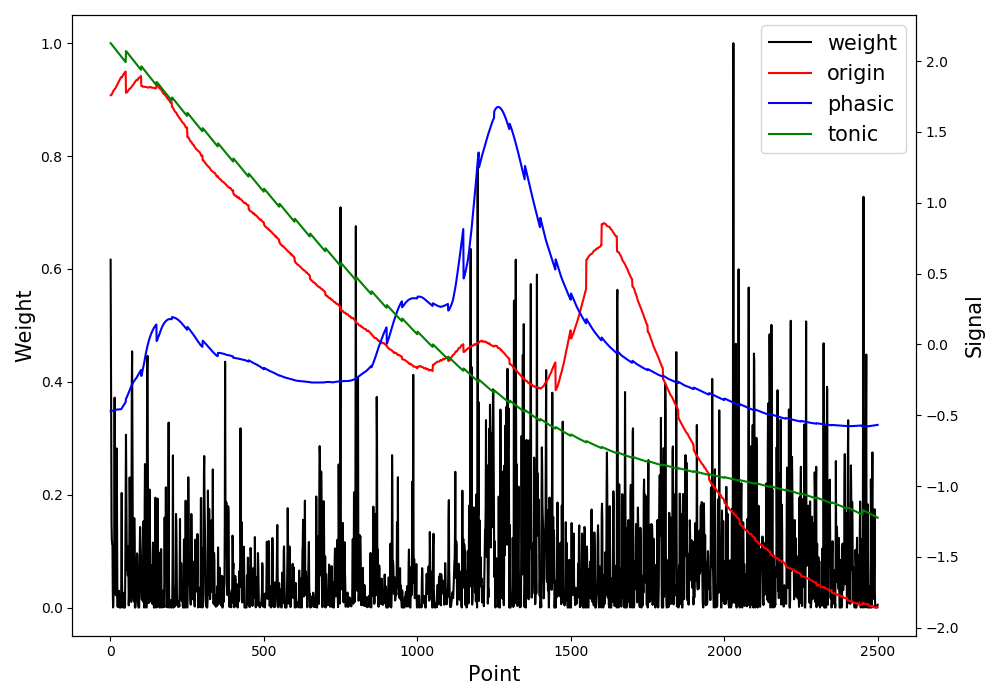}
  }%
  \subfloat[Temporal attention]{
  \includegraphics[width=4.6cm]{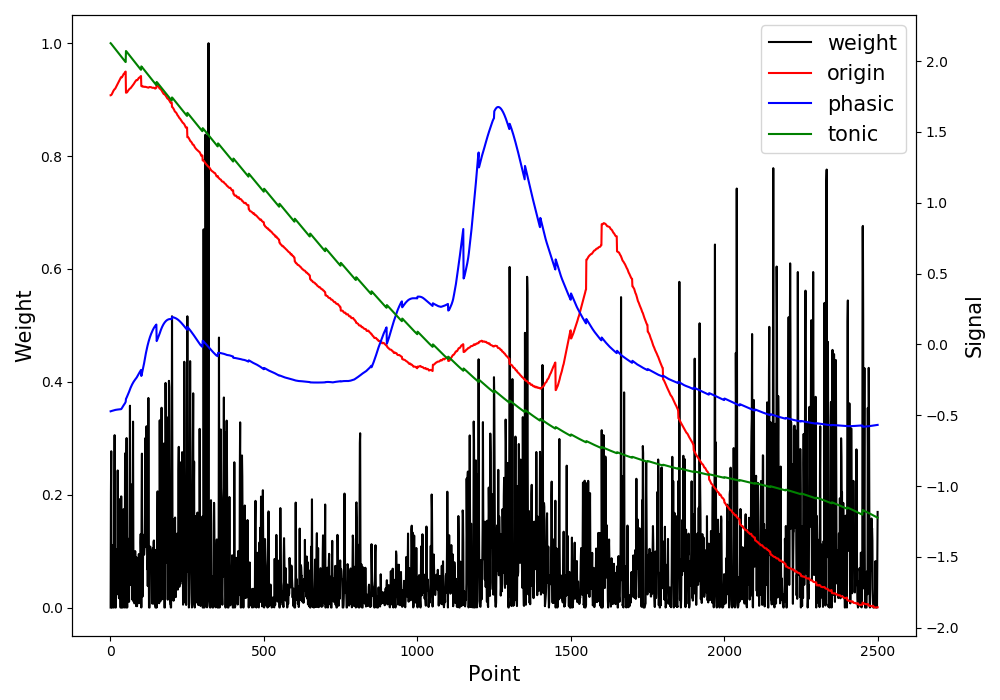}
  }%
  \subfloat[Channel-temporal attention]{
  \includegraphics[width=4.6cm]{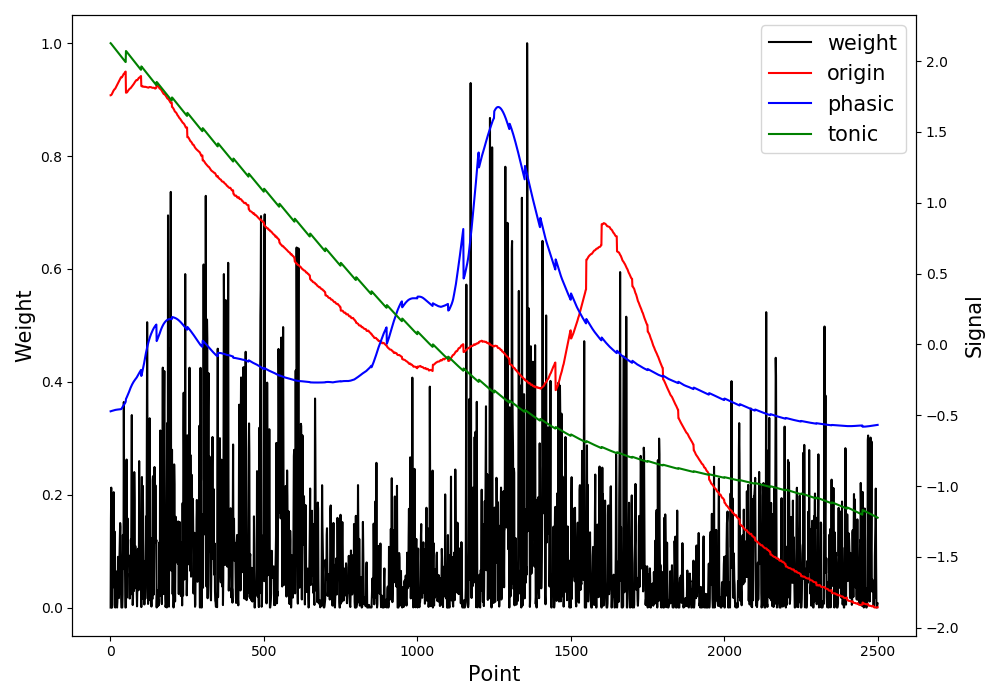}
  }%
\caption{The visualization of the curves of three signals and the column of attention weight in arousal recognition with the random subject (Subject-ID: 100184, Music-ID: 1, fold: 10). The red, blue, and green curves note the origin, phasic and tonic signals, respectively. The black columns represent the inverse mapping of attention weight from the output characteristics of attention submodules to 3-channel input.}
\label{ac}
\end{figure}

\begin{figure}
\centering
  \subfloat[Channel attention]{
  \includegraphics[width=10cm]{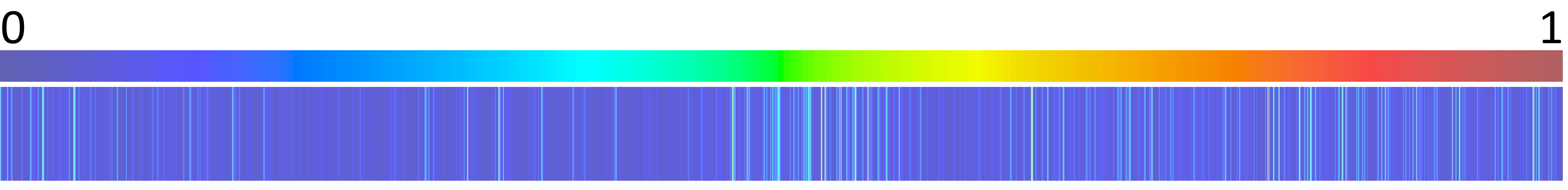}
  }%

    \subfloat[Temporal attention]{
  \includegraphics[width=10cm]{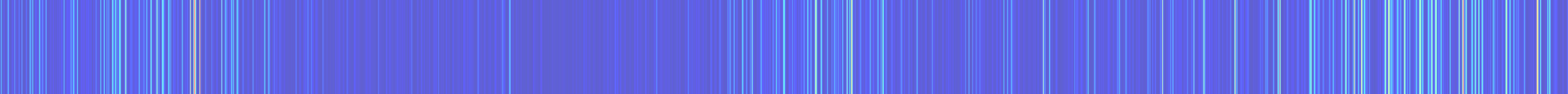}
  }%

    \subfloat[Channel-temporal attention]{
  \includegraphics[width=10cm]{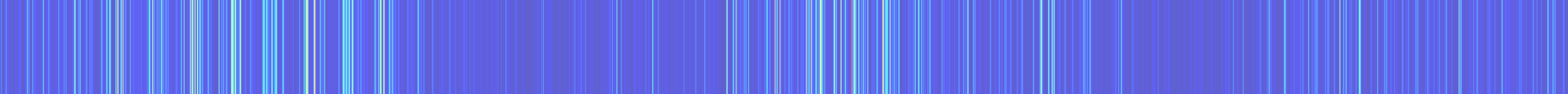}
  }%
\caption{The arousal spectrogram visualization. The spectrograms in (a), (b), (c) separately correspond to the attention weights in Fig. \ref{ac}.}
\label{ah}
\end{figure}

Fig.\ref{vc} and Fig.\ref{ac} show the sequences of pre-processed input signal and output attention weights from Grad-CAM. The input of valence and arousal situation (Subject-ID: 100184, Music-ID: 1) are randomly selected. In order to illustrate the corresponding relationship more clearly, the curves the 3-channel signal curves and the attention weights histogram are shown in one chart. Moreover, attention weights are presented in the form of spectrograms in Fig. \ref{vh} and Fig. \ref{ah}. We can see that the attention weights of the channel attention module consider the relationships between different channels. The weight of attention of the temporal attention module focuses more on the temporal distribution. The attention weights of the integrated channel-temporal attention model selectively cover the regions better than other models because the corresponding spectrogram contains the most warm color, and black columns in the chart are densest in important regions. It is also clear that the integrated RTCAN-1D pays more attention to the sharp peaks in 3 curves.

The quantitative experiment results in Section \ref{4.5.5} and qualitative visualization comparisons in this subsection can show that the channel and temporal attention modules guide the proposed RTCAN-1D to improve the performance of the network.

\begin{table}
\caption{Comparison with previous works in AMIGOS, DEAP, and PMEmo datasets.}
\captionsetup{width=.9\textwidth}
\label{sta}
\centering
\begin{tabular}{ccccccc}
\toprule
\textbf{Model}&\textbf{Dataset}&\textbf{Subjects}&\textbf{Signal}&\textbf{Val}&\textbf{Ave-Acc}&\textbf{F1-Score}\\

\midrule
SVM~\cite{correa2018amigos}&AMIGOS&40&EDA&10-fold&-&\makecell[c]{V: 52.80\% \\A: 54.10\%}\\
DCNN~\cite{santamaria2018using}&AMIGOS&40&EDA&-&\makecell[c]{V: 75.00\% \\A: 71.00\%}&\makecell[c]{V: 71.00\% \\A: 67.00\%}\\
DCNN~\cite{santamaria2018using}&AMIGOS&40&ECG+EDA&-&\makecell[c]{V: 75.00\% \\A: 76.00\%}&-\\
\textbf{RTCAG-1D}&AMIGOS&40&EDA&10-fold&\makecell[c]{\textbf{V: 75.87\%} \\\textbf{A: 78.20\%} }&\makecell[c]{\textbf{V: 74.48\%} \\\textbf{A: 77.56\%}}\\

\midrule
SVM~\cite{koelstra2011deap}&DEAP&32&\makecell[c]{EEG,EOG,EMG\\,GSR,TMP,BVP,RSP}&10-fold&\makecell[c]{V: 62.70\% \\A: 57.00\%}&-\\
MLP~\cite{SHARMA2019324}&DEAP&32&EDA&10-fold&\makecell[c]{V: 69.80\% \\A: 79.00\%}&-\\
CNN~\cite{2019A}&DEAP&32&EDA&10-fold&\makecell[c]{V: 82.00\% \\A: 82.00\%}&\makecell[c]{\textbf{V: 83.00\%}\\A: 83.00\%}\\
RTCAG-1D&DEAP&32&EDA&10-fold&\makecell[c]{V: 77.15\% \\A: 83.42\%}&\makecell[c]{V: 79.18\%\\\textbf{A: 83.77\%}}\\
\textbf{Alexnet-2D}~\cite{lin2017deep}&DEAP&32&\makecell[c]{EEG,EOG,EMG\\,GSR,TMP,BVP,RSP}
&10-fold&\makecell[c]{\textbf{V: 87.30\%} \\ \textbf{A: 85.50\%}}&\makecell[c]{V: 78.24\%\\A: 80.06\%}\\

\midrule
Res-SIN~\cite{yin2019user}&PMEmo&457&EDA&10-fold&\makecell[c]{V: 55.92\%\\A: 57.24\% }&\makecell[c]{V: 58.83\%\\A: 60.12\%}\\
RTCAN-1D&PMEmo&457&EDA&10-fold&\makecell[c]{V: 64.52\%\\A: 65.19\% }&\makecell[c]{V: 63.31\%\\A: 65.84\%}\\
Res-SIN~\cite{yin2019user}&PMEmo&457&EDA+music&10-fold&\makecell[c]{V: 73.43\%\\A: 73.65\% }&\makecell[c]{V: 77.54\%\\A: 78.56\%}\\
\textbf{RTCAN-1D}&PMEmo&457&EDA+music&10-fold&\makecell[c]{\textbf{V: 79.68\%}\\ \textbf{A: 83.76\%} }&\makecell[c]{\textbf{V: 82.45\%}\\ \textbf{A: 86.12\%}}\\
\bottomrule
\end{tabular}
\end{table}

\subsection{Comparison with previous methods}
To validate our proposed network's superiority, we provide extensive comparisons in three emotion datasets. The single EDA signals from the small-scale DEAP and AMIGOS datasets are used to compare our RTCAG with SOTA EDA-based methods. And large-scale PMEmo with music and EDA signals is applied to prove the effectiveness of our multimodal fusion. Those SOTA models include both the hand-crafted feature-based conventional classifiers and deep learning-based frameworks.

For single EDA, previous works applied various models to classify the hand-craft features such as Support Vector Machine (SVM)~\cite{correa2018amigos,koelstra2011deap}, Multi-Layer Perceptron (MLP)~\cite{SHARMA2019324}. The CNN-based methods~\cite{lin2017deep,santamaria2018using,yin2019user} with auto feature extraction were also proposed. We notice that the EDA decomposition is gradually being used for emotion recognition task~\cite{SHARMA2019324}. And our previous work~\cite{yin2019user} has also combined CvxEDA~\cite{greco2015cvxeda} and ResNet~\cite{he2016deep} for emotion recognition. Recently, Machot \textit{et al.}~\cite{2019A} trained a 3-layer SOTA CNN using a grid search technique to get 82\% accuracy on the subject-independent model. It should be noted that there is not much EDA-based emotional recognition in previous works. For example, there is no baseline use of EDA signals for valence and arousal classification in DEAP dataset. Hence, for a comprehensive evaluation, the comparisons are not limited to the EDA-based model. Other multimodal psychological signal-based methods (EEG,ECG) make use of traditional classifiers ($e.g.,$ SVM~\cite{koelstra2011deap}) and deep learning model ($e.g.,$ DCNN~\cite{lin2017deep}). Table \ref{sta} lists the comparisons between our model and previous studies. Our RTCAG outperforms existing EDA-based SOTA methods. It should be noted that for binary recognition in DEAP, the evaluation results of our model in valence dimension are a little lower than 3 layers CNN model~\cite{2019A} because they directly mapped the scales ($1\sim 9$) into 2 levels. For example, the valence scale of $1\sim 5$ was mapped to “negative” and $6\sim 9$ to “positive”, respectively. The arousal scale of $6\sim 9$ was mapped to “passive” and $6\sim 9$ to “active”, respectively~\cite{2019A}. The direct value annotation labeling of Machot \textit{et al.}~\cite{2019A} simplified the recognition problem without considering the subject emotion threshold. Moreover, Alexnet-2D~\cite{lin2017deep} has fused 7 physiological signals, \emph{i.e.} EEG, EOG, EMG, GSR, TMP, BVP, RSP, it's natural that our RTCAG-1D only with EDA signals cannot outperform Alexnet-2D~\cite{lin2017deep} but the competitive results can also prove the effectiveness of our model.

For multimodal solution, the better performance of mixed-method multiphysiological signals also indicates the multi-feature fusion as mixed inputs contain more affective messages than a single input if acquisition equipments are sufficient. However, considering the user-friendly applications for reliability and hard acquisition, our RTCAG does not involve more physiological signals but adds the music features. With the effective temporal and channel attention mechanism, the representation ability of our well-designed RTCAG has improved to mine more useful EDA features. Hence, the performance of RTCAN-1D gets a significant improvement than our previous Res-SIN~\cite{yin2019user} without changing music features. All the results in three datasets validate the superiority of our RTCAN-1D.

\section{Conclusion and Discussion}
The individual differences in the large-scale emotion recognition task is a tough problem, which causes the performance drop of user independent methods. The single input music or EDA signal cannot alleviate this problem. Towards this, we make an attempt with multimodal fusion. In this work we propose an end-to-end framework called the 1-dimensional residual temporal and channel attention network for user-independent emotional recognition.
As the first attempt, our proposed RTCAN-1D separately uses the temporal-channel attention modules to mine the long-range temporal correlations and channel-wise relationships to reweight the distribution of the shallow features. Moreover, in large-scale PMEmo dataset, the fusion of emotion benchmarks from external music stimulation and the individual specificity from EDA also significantly improves the performance of our model.
Extensive experiments in three popular datasets validate the effectiveness of our model. In future work, the fusion of external static features and traditional physiological signals will be an efficient way for large-scale affective computing. The attention mechanism can also further improve the ability of CNN's end-to-end frameworks. We believe the research of EDA-based emotion recognition will promote a lot of practical applications.


\bibliographystyle{ACM-Reference-Format}
\bibliography{sample-base}

\end{document}